\documentstyle[12pt, epsfig]{article}     
     
\begin{document}     
     
\vspace*{-2cm} \renewcommand{\thefootnote}{\fnsymbol{footnote}}     
\begin{flushright}     
hep-ph/0012157\\     
PSI PR-00-18\\ 
December 2000\\     
\end{flushright}     
\vskip 65pt     
     
\begin{center}     
{\Large {\bf Resummation of Yukawa enhanced and subleading Sudakov logarithms in longitudinal
gauge boson and Higgs production
}}\\[0pt]     
\vspace{1.2cm}    
{Michael Melles\footnote{{\bf     
Michael.Melles@psi.ch}} }\\     
\vspace{10pt}   
  
{Paul Scherrer Institute (PSI), CH-5232 Villigen, Switzerland. }  
\end{center}

\vspace{60pt}     
\begin{abstract}     
Future colliders will probe the electroweak theory at energies much larger than the
gauge boson masses. Large double (DL) and single (SL) logarithmic virtual 
electroweak Sudakov corrections 
lead to significant effects for observable cross sections. Recently, leading and subleading
universal corrections for external fermions and transverse gauge boson lines were resummed by employing
the infrared evolution equation method. The results were confirmed at the DL level by explicit
two loop calculations with the physical Standard Model (SM) fields. 
Also for longitudinal degrees of freedom the
approach was utilized for DL-corrections 
via the Goldstone boson equivalence theorem. In all cases, the electroweak Sudakov logarithms
exponentiate. In this paper we extend the same approach to both Yukawa enhanced as well as
subleading Sudakov corrections to longitudinal gauge boson and Higgs production. We use
virtual contributions to splitting functions of the appropriate Goldstone bosons
in the high energy regime and find that all universal subleading terms exponentiate. The approach is
verified by employing a non-Abelian version of Gribov's factorization theorem and by
explicit comparison with existing one loop calculations. 
As a side result, we obtain also all top-Yukawa enhanced subleading logarithms for chiral fermion
production at high energies to all orders.
In all cases, the size of the subleading contributions at the two loop level is non-negligible
in the context of precision measurements at future linear colliders.
\end{abstract}

\vskip12pt     
     
\setcounter{footnote}{0} \renewcommand{\thefootnote}{\arabic{footnote}}     
     
\vfill     
\clearpage     
\setcounter{page}{1} \pagestyle{plain}     
     
\section{Introduction}     

The high energy behavior of the Standard Model (SM) will become increasingly important
at future colliders investigating the origin of electroweak symmetry breaking. 
At the expected level of precision required to disentangle new physics effects from the SM
in the ${\cal O} \left( \leq 1 \% \right)$ regime, higher order {\it electroweak} radiative
corrections cannot be ignored at energies in the TeV-range. The largest contribution
is contained in electroweak double logarithms (DL) of the Sudakov type and a comprehensive
treatment of those corrections is given in Ref. \cite{flmm} to all orders.
The effects of the mass-gap between the photon and Z-boson has been 
considered in recent publications \cite{m2,bw} since spontaneously broken gauge theories lead
to the exchange of massive gauge bosons. In general one expects the
SM to be in the unbroken phase at high energies.
There are, however, some important differences of the electroweak theory with respect to an unbroken
gauge theory. Since the physical cutoff of the massive gauge bosons is the weak scale $M\equiv 
M_{\rm W} \sim M_{\rm Z} \sim M_{\rm H}$,
pure virtual corrections lead to physical cross sections depending on the infrared ``cutoff''.
Only the photon needs to be treated in
a semi-inclusive way.
Additional complications arise due to the mixing involved to make the mass eigenstates and the fact
that at high energies, the longitudinal degrees of freedom are not suppressed.
Furthermore, since the asymptotic states are not group singlets, it is expected
that fully inclusive cross sections contain Bloch-Nordsieck violating electroweak corrections 
\cite{ccc}.

It has by now been established that the
exponentiation of the electroweak Sudakov DL calculated in Ref.
\cite{flmm} via the infrared evolution equation method \cite{kl} with the fields of 
the unbroken phase is indeed reproduced by explicit
two loop calculations with the physical SM fields \cite{m2,bw,hkk}. One also understands now
the origin of previous disagreements. The results of Ref. \cite{kp}, based on fully
inclusive cross sections in the photon, is simply not gauge invariant as already
pointed out in Ref. \cite{flmm}. The factorization used in Ref. \cite{cc} is based
on QCD and only takes into account contributions from ladder diagrams. In the electroweak
theory, the three boson vertices, however, do not simply cancel the corresponding
group factors of the crossed ladder diagrams (as is the case in QCD) and thus, infrared
singular terms survive for left handed fermions (right handed ones are effectively
Abelian) in the calculation of Ref. \cite{cc}. The infrared evolution equation method does
not encounter any such problems since all contributing diagrams are automatically
taken into account by determining the kernel of the equation in the effective regime
above and below the weak scale $M$. 
It is then possible to calculate corrections in the effective high energy theory
in each case yielding the same result as calculations in the physical basis.
Thus, the
mass gap between the Z-boson and the photon can be included in a natural way
with proper matching conditions at the scale $M$.
For longitudinally polarized gauge bosons it was shown in Ref. \cite{m1} that
the DL kernel can be obtained from the Goldstone boson equivalence theorem.

The picture that emerges has a clear physical interpretation. At high energies,
where particle masses can be neglected,
the effective theory is given by an unbroken $SU(2) \times U(1)$ theory for
fermions and transversely polarized gauge bosons, and by the equivalence theorem
for longitudinally polarized gauge bosons. The contribution from soft photons
and collinear terms below the weak scale is determined by QED (including mass
terms in the corresponding logarithms).
This approach was utilized in Ref. \cite{m1} to obtain the subleading (SL) universal
terms to all orders for external fermion lines (up to Yukawa enhanced terms)
and transversely polarized gauge
bosons. Universal, i.e. process independent,
are those terms which through Ward identities can be related
to external lines and at high energies are given by the contributions to
the virtual splitting functions (see Ref. \cite{m1}). In addition, there
are non-universal angular terms of the type $\log \frac{u}{t} \log \frac{s}{M^2}$
which can be important and should be included at least at one loop. In general, these
terms cannot be resummed, they are process dependent
and don't factorize with respect to the Born amplitude. At one loop, however,
there is a general method for calculating such terms \cite{dp} and in practice
at most a two loop approach to subleading logarithmic accuracy would be needed. 

An important aspect of resumming universal terms is given by the fact that there
is a partial cancellation between the DL and SL corrections at energies around a
TeV, thus enabling one to see how reliable a DL analysis for a given process
really is. In addition, these are predictions of universal terms which can always
by used to check higher order calculations which in the electroweak theory are
extremely involved due to the number of mass terms and diagrams contributing.
It is also conceptually important for a theoretical understanding of the infrared
behavior of the SM.
By comparing the subleading universal terms with existing one-loop calculations,
we gain further evidence for the overall method employed, in particular
when it comes to understanding differences between unbroken and broken gauge theories.

From a phenomenological point of view, the corrections to longitudinal
gauge boson production are important in case of a strongly interacting
$W^\pm$ sector without a fundamental Higgs boson. Our perturbative approach
would of course break down in that case, however, it is important to know
the precise form of the deviation of the new dynamics from the SM in the
TeV range in order to understand the new physics behind the electroweak
breaking sector.
Corrections to Higgs bosons are important for a precise measurement of
the Yukawa couplings at high energies in order to establish the Higgs mechanism. 
At the level of $6-8\%$, these corrections can certainly not be neglected
at 1 TeV for determinations of the top-Yukawa coupling at $e^+e^-$ 
colliders \cite{bdr}. 

In this paper, we complete the all orders resummation of all SL universal
Sudakov corrections to the SM. While in Ref. \cite{m1} have restricted ourselves
to calculating terms analogous to QCD, we now consider terms typical for broken
gauge theories. These are in particular longitudinal degrees of freedom, 
processes with external Higgs bosons and Yukawa enhanced logarithmic corrections.

In section \ref{sec:et} we discuss how SL contributions are obtained in the
scalar sector and the application of the equivalence theorem. In section \ref{sec:sf}
we give results for the virtual contributions to splitting functions involving
Goldstone and Higgs bosons fulfilling evolution equations analogous to the 
Altarelli-Parisi equations. The correctness of this approach to the SL level
is verified in section
\ref{sec:gt} by employing a non-Abelian version of Gribov's bremsstrahlung theorem
to processes involving Yukawa enhanced contributions. A similar approach can
be utilized to verify analogous corrections for chiral fermions ($b_{\rm L},
t_{\rm L}, t_{\rm R})$ in section \ref{sec:cf}. Semi-inclusive cross sections
for physical observables are given in section \ref{sec:si} and we
compare our results with existing one loop calculations in section \ref{sec:ol}.
We discuss the size of the corrections obtained in section \ref{sec:dis}
and present our conclusions in section \ref{sec:co}.

\section{The effective high energy theory} \label{sec:et}

In the following we discuss the corrections in a non-Abelian theory with scalar external
``quarks'', i.e. external scalar bosons charged under the unbroken gauge group.
The physical picture is that at high energies, we can use this effective theory via
the Goldstone boson equivalence theorem to describe the longitudinal degrees of freedom.
The latter will be discussed in detail in section \ref{sec:gst}. An additional complication 
is given by the presence of Yukawa enhanced logarithmic corrections which are a novel 
feature of theories
with spontaneously broken gauge symmetry. These terms are discussed in
section \ref{sec:sf}. We begin with a discussion of 
scalar (massless) QCD at high energies. 

\subsection{Scalar QCD} \label{sec:sQCD}

In this section we are interested in the collinear
corrections to external scalars in a non-Abelian gauge theory
at the subleading level when the external legs are taken on the mass shell.
According to the discussion in Ref. \cite{m1} we have to calculate  
terms contributing to anomalous scaling violations. For purely virtual corrections
the invariant matrix element fulfills the following differential equation
for massless scalar quarks and all invariants $2p_jp_l \sim s$ large compared
to an infrared cutoff $\mu$ and denoting $t=\log \frac{s}{\mu^2}$:
\begin{eqnarray}
&& \left( \frac{\partial}{\partial t} + \beta^{\rm sQCD} \frac{\partial}{\partial g_s} + n_g \left(
\Gamma_g(t)- \frac{1}{2} \frac{\alpha_s}{\pi} \beta^{\rm sQCD}_0 \right) + n_s
\left( \Gamma_s(t) + \frac{1}{2} \gamma_{s\overline{s}} \right) \right) \nonumber \\
&& \times {\cal M} (p_1,...,p_n,g_s,\mu^2) =0 \label{eq:mrg}
\end{eqnarray}
to the order we are working here and
where ${\cal M}(p_1,...,p_n,g_s,\mu)$ is taken on the mass-shell.
The $\Gamma (t)$ are infrared singular anomalous dimensions leading to DL corrections
given in Ref. \cite{m1}
while the gluon and scalar quark anomalous dimensions describe SL contributions.
At higher orders, the subleading RG corrections can be incorporated by including a running
coupling in each loop \cite{ms2}.
An additional non-mass suppressed term occurs in the electroweak theory
in the case of four scalar scattering
amplitudes (such as $\phi^+\phi^- \longrightarrow \phi^+\phi^-$) due to the mass ratios in
the coupling ($\lambda \sim \frac{m_{\rm H}^2}{M^2}$) at the Born level. In that case one has to
renormalize not only the gauge couplings but also the respective scalar couplings
at the one loop level
\cite{dp}. At higher orders, mass renormalization corrections are subsubleading. 

The universality of the corrections follows from
the spin independence of the factorization theorems \cite{facth} and thus, from the universality
of the splitting functions in scalar QCD\footnote{I would like to thank J.~Collins for
helpful discussions on this point.}. Such an equation is of course not physical for a theory
with massless gauge bosons, however, in the electroweak theory purely virtual corrections
can lead to physical cross sections. Only real soft photon emission needs to be included.

We are interested in determining the scalar quark anomalous dimension
$\gamma_{s\overline{s}}$ which gives rise to logarithmic corrections due to scaling
violations from the classical case. The factor of $\frac{1}{2}$ in Eq. (\ref{eq:mrg}) originates
from the fact that it is written for each external leg separately.

\begin{figure}
\centering
\epsfig{file=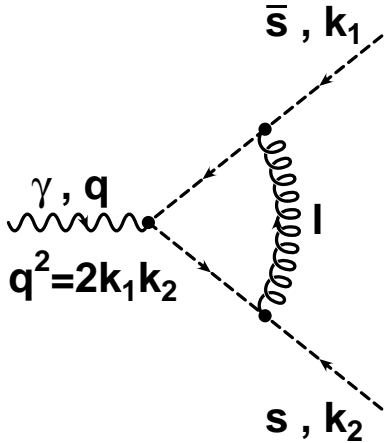,width=5cm}
\caption{A Feynman diagram determining the DL and SL contributions to scalar quarks
in the on-shell scheme. In the massless theory there are scaling violations from loop
corrections which can be described by anomalous dimensions.}
\label{fig:sQCD}
\end{figure}

Thus we must at one loop calculate the corrections depicted in
Fig. \ref{fig:sQCD} with the corresponding Feynman rules for scalar QCD: 
\begin{eqnarray}
{\cal A}^{(1)}_\nu
&=& - T^b C_F g_s^3 \int \frac{d^nl}{(2 \pi)^n} \frac{
(-4k_1k_2+2l(k_1-k_2)+l^2)(2l+k_1-k_2)_\nu }{
(l^2-\lambda^2+i \varepsilon)((l+k_1)^2+i \varepsilon) ((l-k_2)^2+i \varepsilon)}
\end{eqnarray}
This is in complete analogy to the situation in QCD \cite{ap,a}.
Requiring that the
self energy corrections vanish on the mass shell and the on-shell vertex for zero momentum
transfer\footnote{We can perform this renormalization here since the non-Abelian components
don't enter for the scalar quark anomalous dimension. The corresponding counterterm includes
automatically the wave function renormalization contribution.} we find the following one loop result:
\begin{equation}
{\cal M}^{(1)} = {\cal M}^{\rm Born} \left\{ 1 +C_F \frac{g_s^2}{8 \pi^2} \left( - \frac{1}{2}
\log^2 \frac{s}{\lambda^2} + 2 \log \frac{s}{\lambda^2} - 8 + 2 \frac{\pi^2}{3} \right) \right\}
\label{eq:sqv}
\end{equation}
where ${\cal M}^{\rm Born}=-ig_s T^b (k_1-k_2)_\nu $.
It is important to point out that both leading and subleading logarithmic corrections
factorize with respect to the same group factor. The reproduction of this fact in the
electroweak theory when compared to exact calculations with the physical fields is crucial
in establishing the overall correctness of our approach since we must obtain the factorized
form with respect to the effective high energy theory.
The finite terms in Eq. (\ref{eq:sqv}) 
are of course irrelevant to the discussion here and the infrared divergent
soft and collinear terms were regulated using a fictitious gluon mass term. The difference to
the QED result \cite{mth} for the on-shell form factor
\begin{equation}
{\cal M}^{(1)}_{\rm QED} = {\cal M}^{\rm Born}_{\rm QED} \left\{ 1 +\frac{e^2}{8 \pi^2} \left( - \frac{1}{2}
\log^2 \frac{s}{\lambda^2} + \frac{3}{2} \log \frac{s}{\lambda^2} - 2 + 2 \frac{\pi^2}{3} \right) \right\}
\label{eq:QEDv}
\end{equation}
is (besides the coupling) mainly present in the different collinear divergent subleading term.
This term differs due to the different spin of the particle emitting the gauge boson.
Here ${\cal M}^{\rm Born}_{\rm QED} = -i e \langle k_1, \tau | \gamma_\nu | k_2, \tau \rangle$ 
as usual and replacing $e^2$ with $C_F g^2_s$ we obtain the QCD result from Eq. 
(\ref{eq:QEDv}). 
Scaling violations for S-matrix elements
can be described by calculating the anomalous dimension of the relevant
gauge invariant operators. This is due to the fact that for massless theories there is a one to one
correspondence between high and low energy scaling \cite{pqz,p}.
Thus for the subleading scaling violations
only the regions of large loop integration $l$ in Eq. (\ref{eq:sqv})
are relevant here (the double logarithms lead to infrared singular anomalous
dimensions \cite{kr}) and the corresponding
anomalous dimension can be read off from the subleading logarithmic term:
\begin{equation}
\gamma_{s {\overline s}} = \frac{\partial}{ \partial \log {\overline \mu}} (-\delta_{
s {\overline s}} + \delta_s)=- C_F \frac{\alpha}{\pi} \label{eq:sad}
\end{equation}
where $\delta_{s {\overline s}}$ denotes the counterterm from the diagram depicted in Fig.
\ref{fig:sQCD}, while $\delta_s$ corresponds to the wave function renormalization counterterm
of the scalar quark. The sum in Eq. \ref{eq:sad} is gauge independent.
Since the factorization theorems of Refs. \cite{facth} do not depend on the spin of the quark, we can
resum the leading and subleading virtual logarithmic corrections by using the Altarelli-Parisi
equations. To this end we must formulate the above results in terms of the language of the
splitting functions for a massless scalar quark. This will be done in the next section. First,
however, we are going to discuss the scalar high energy sector in the Standard Model. In particular,
there are additional corrections of the Yukawa type which need to be discussed, for
which there is no analogue in unbroken gauge theories.

\subsection{The equivalence theorem} \label{sec:gst}

At high energies, the longitudinal polarization states can be described with the polarization
vector 
\begin{equation}
e_L^\nu (k)=k^\nu/M + {\cal O}(M/E_k) \label{eq:lpv} 
\end{equation}
The connection between S-matrix elements and Goldstone bosons
is provided by the equivalence theorem \cite{gb}. It states that at tree level for S-matrix
elements for longitudinal
bosons at the high
energy limit $M^2/s\longrightarrow 0$ can be expressed through matrix elements
involving their associated
would be Goldstone bosons. We write schematically in case of a single gauge boson:
\begin{eqnarray}
{\cal M}(W^\pm_{L}, \psi_{{\rm phys}}) &=& {\cal M}(\phi^\pm, \psi_{{\rm phys}}
) + {\cal O} \left(
\frac{M_{\rm w}}{\sqrt{s}} \right)
\label{eq:wet} \\
{\cal M}(Z_{L}, \psi_{{\rm phys}}) &=& i {\cal M}(\chi, \psi_{{\rm phys}}) + {\cal
O} \left(
\frac{M_{\rm z}}{\sqrt{s}} \right)
\label{eq:zet}
\end{eqnarray}
The problem with this statement of the equivalence theorem is that it holds only
at tree level
\cite{yy,bs}. For
calculations at higher orders,
additional terms enter which change Eqs. (\ref{eq:wet}) and (\ref{eq:zet}).

Because of the gauge invariance of the physical theory and the associated BRST
invariance, a modified
version of Eqs. (\ref{eq:wet}) and (\ref{eq:zet}) can be derived \cite{yy} which
reads
\begin{eqnarray}
k^\nu {\cal M}(W^\pm_{\nu}(k), \psi_{{\rm phys}}) &=& C_{\rm w} M_{\rm w} {\cal
 M}(\phi^\pm (k),
 \psi_{{\rm phys}}) + {\cal O} \left( \frac{M_{\rm w}}{\sqrt{s}} \right) \label{
 eq:wetp} \\
 k^\nu{\cal M}(Z_{\nu}(k), \psi_{{\rm phys}}) &=& i C_{\rm z} M_{\rm z} {\cal M}(
 \chi (k), \psi_{{\rm phys}})
 + {\cal O} \left( \frac{M_{\rm z}}{\sqrt{s}} \right) \label{eq:zetp}
 \end{eqnarray}
 where the multiplicative factors $C_{\rm w}$ and $C_{\rm z}$ depend only on wave
 function renormalization
 constants and mass counterterms. Thus, using the form of the longitudinal 
 polarization vector of
 Eq. (\ref{eq:lpv}) we can write
 \begin{eqnarray}
 {\cal M}(W^\pm_{L}(k), \psi_{{\rm phys}}) &=&  C_{\rm w} {\cal M}(\phi^\pm (k),
 \psi_{{\rm phys}}) + {\cal O} \left( \frac{M_{\rm w}}{\sqrt{s}} \right) 
 \label{eq:wgs} \\
 {\cal M}(Z_{L}(k), \psi_{{\rm phys}}) &=& i C_{\rm z} {\cal M}(\chi (k), \psi_{{
 \rm phys}})
 + {\cal O} \left( \frac{M_{\rm z}}{\sqrt{s}} \right) \label{eq:zgs}
 \end{eqnarray}
 We see that in principle,
 there are logarithmic loop corrections to the tree level equivalence theorem.
 The important point in our approach, however, is that the correction coefficients
 are not functions of the energy variable $s$: 
 \begin{equation}
 C_{\rm w}= C_{\rm w} ( \overline{\mu}, M, g, g^\prime) \;\;, \; 
 C_{\rm z}= C_{\rm z} ( \overline{\mu}, M, g, g^\prime) \label{eq:etcc}
 \end{equation}
 \begin{figure}
 \centering
 \epsfig{file=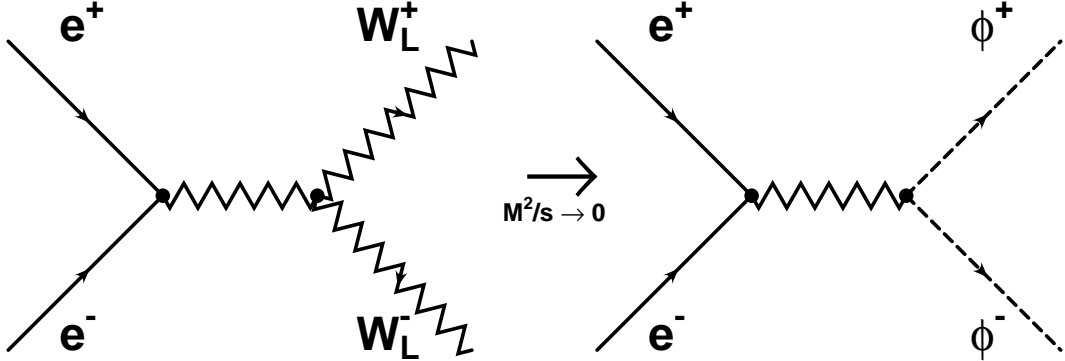,width=14cm}
 \caption{The pictorial Goldstone boson equivalence theorem for $W$-pair production in 
 $e^+e^-$ collisions.
 The correct DL-asymptotics for longitudinally polarized bosons
 are obtained by using the quantum numbers of the charged would be Goldstone scalars 
 at high energies.}
 \label{fig:gsb}
 \end{figure}
 The pictorial form of the Goldstone boson equivalence theorem is depicted in Fig. \ref{fig:gsb}
 for longitudinal $W$-boson production at a linear $e^+e^-$ collider.
 In the following we denote the logarithmic variable $t\equiv \log \frac{s}{\mu^2}$, where $\mu$ is
 a cutoff on the transverse part of the exchanged virtual momenta $k$ of all involved particles, i.e.
 \begin{equation}
 \mu^2 \leq \mbox{\boldmath $k$}^2_\perp \equiv \min (2 (kp_l)(kp_j)/(p_lp_j))
 \label{eq:kpdef}
 \end{equation}
 for all $j \neq l$. The non-renormalization group part of the evolution equation at high
 energies is given on the invariant matrix element level by \cite{m1}:
\begin{equation}
\frac{\partial}{\partial t} {\cal M}(L(k), \psi_{{\rm phys}}) = K (t)
{\cal M}(L(k), \psi_{{\rm phys}}) \label{eq:leveq}
\end{equation}
and thus, after inserting Eqs. (\ref{eq:wgs}), (\ref{eq:zgs}) we find that the same evolution equation
also holds for ${\cal M}(\phi(k), \psi_{{\rm phys}})$. The notation here is $L=\{W_L^\pm, Z_L \}$
and $\phi=\{ \phi^\pm, \chi\}$, respectively.
Thus, the $\log \frac{s}{\mu^2}$ dependence in our approach is unrelated to the corrections
to the equivalence theorem, and in general, is unrelated to two point functions in a 
covariant gauge at high energies where masses can be neglected. 
This is a consequence of the physical on-shell renormalization scheme
where the $\overline{\rm MS}$ renormalization scale parameter ${\overline \mu} \sim M$.
Physically, this result can be understood by interpreting the correction terms $C_{\rm w}$
and $C_{\rm z}$ as corrections required by the gauge invariance of the theory in order
to obtain the correct renormalization group asymptotics of the physical Standard Model fields.
Thus, their origin is not related to Sudakov corrections.
In other words,
the results from the previous section should be applicable to the subleading scalar sector
in the electroweak theory regarding a non-Abelian scalar gauge theory as the effective 
description in this range. The only additional complication in the Standard Model is
the presence of subleading Yukawa enhanced logarithmic corrections which will be discussed
below. It is also worth noticing, that at one loop, the authors of Ref. \cite{dp} obtain
the same result for the contributions from the terms of Eq. (\ref{eq:etcc}). In their approach,
where all mass-singular terms are identified and the renormalization scale ${\overline \mu}
=\sqrt{s}$,
these terms are canceled by additional corrections from mass and wave function counterterms.
At higher orders it is then clear that corrections from two point functions are subsubleading
in a covariant gauge.

\section{Subleading corrections from splitting functions} \label{sec:sf}

\begin{figure}
\centering
\epsfig{file=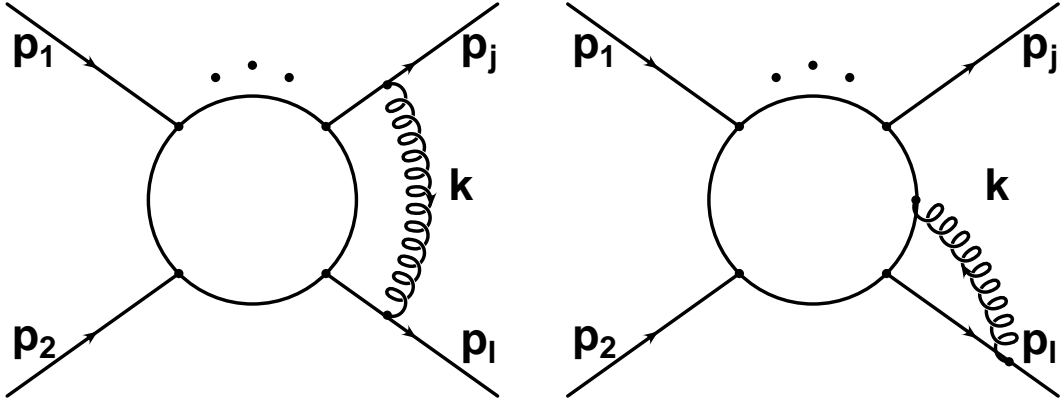,width=14cm}
\caption{Feynman diagrams contributing to the infrared evolution
equation (\ref{eq:mrg}) for a process with $n$ external scalar quarks. In a general
covariant gauge the
virtual gluon with the smallest value of ${\mbox{\boldmath $k$}}_{\perp}$ is attached to
different external lines. The inner scattering amplitude is assumed to be
on the mass shell.} \label{fig:nll}
\end{figure}

In an axial gauge, collinear logarithms are related to corrections
on a particular external leg depending on the choice of the four vector $n_\nu$ \cite{col}.
In a covariant gauge, the sum over all possible
insertions shown in Fig. \ref{fig:nll} with all invariants large ($\sim s$)
is reduced to a sum over all $n$-external legs due to Ward identities.
Overall, these
corrections factorize with respect to the Born amplitude.
We can therefore adopt the strategy to extract the gauge invariant contribution
from the external
line corrections on the invariant matrix element at the subleading level.
In Ref. \cite{m1} we showed that in the high energy regime, subleading logarithmic
corrections in massless theories are of collinear or RG origin.
This is important since it allows us to use the Altarelli-Parisi
approach to calculate the subleading contribution to the evolution kernel of Eq.
(\ref{eq:leveq}).
We are here only concerned with virtual corrections and use the universality of
the splitting
functions to calculate the subleading terms. For longitudinal degrees of freedom
we have shown that to logarithmic accuracy the electroweak theory can be described
by scalar Goldstone bosons via the equivalence theorem. Thus at high energies, the
effective theory is analogous to scalar QCD\footnote{Although Yukawa terms are
not present in QCD with scalar quarks, we will show in the next section that
at subleading level the Yukawa terms can be treated as an additional term
in the Altarelli-Parisi splitting function for Goldstone bosons.}.
For this purpose we use the virtual
gauge boson
contributions to the splitting functions
$P^V_{\phi^\pm \phi^\pm}(z)$, $P^V_{\chi \chi}(z)$ and $P^V_{\rm HH}(z)$ 
describing the probability to emit a soft and/or
collinear virtual
particle with energy fraction $z$ of the original external line four momentum.
The infinite momentum frame corresponds to the Sudakov parametrization with 
lightlike vectors.
In general, the splitting functions $P_{\rm BA}$
describe the probability of finding a particle $B$ inside a particle $A$
with fraction $z$
of the longitudinal momentum of $A$ with probability
${\cal P}_{\rm BA}$ to first order \cite{ap}:
\begin{equation}
d {\cal P}_{\rm BA}(z)=\frac{\alpha_s}{2\pi} P_{\rm BA} d t
\end{equation}
where the variable $t=\log \frac{s}{\mu^2}$ for our purposes. It then follows 
\cite{ap} that
\begin{equation}
d {\cal P}_{\rm BA}(z)=\frac{\alpha_s}{2\pi} \frac{z(1-z)}{2} \overline{\sum_{spins
}} \frac{|V_{A
\longrightarrow B+C}|^2}{{\mbox{\boldmath$k$}^2_{\perp}}} d \log {\mbox{\boldmath
$k$}^2_{\perp}} \label{eq:dpabres}
\end{equation}
where $V_{A\longrightarrow B+C}$ denotes the elementary vertices and
\begin{equation}
P_{\rm BA}(z)=\frac{z(1-z)}{2} \overline{\sum_{spins}} \frac{|V_{A
\longrightarrow B+C}|^2}{{\mbox{\boldmath$k$}^2_{\perp}}}
\end{equation}
The upper bound on the integral over $d {\mbox{\boldmath$k$}^2_{\perp}}$ in Eq. (\ref{eq:dpabres})
is $s$ and it is thus
directly related to $d t$.
Regulating the virtual infrared divergences with the transverse momentum cutoff as described above,
we find the virtual contributions to the splitting functions for external Goldstone and Higgs
bosons:
\begin{eqnarray}
&& P^V_{\phi^\pm \phi^\pm}(z) = P^V_{\chi \chi}(z) = P^V_{HH}(z) = \nonumber \\
&& \left[ \left( T_i(T_i+1)+  \tan^2 \theta_{\rm w} \left( \frac
{Y_i}{2}
\right)^2 \right) \left( - 2 \log \frac{s}{\mu^2} + 4 \right) 
- \frac{3}{2} \frac{m^2_t}{M^2} \right] \delta(1-z)
\label{eq:pv}
\end{eqnarray}
The functions can be calculated directly from loop corrections to the elementary
processes in analogy to QCD \cite{aem,a,dot} and the logarithmic term corresponds to the
leading kernel of Ref. \cite{m1}.
We introduce virtual distribution functions which include only the effects of loop computations.
These fulfill the Altarelli-Parisi equations\footnote{Note that off diagonal
splitting functions do not contribute to the virtual probabilities to the order
we are working here. In fact, for virtual corrections there is no need to introduce off-diagonal terms
as the corrections factorize with respect to the Born amplitude. The normalization of Eq.
(\ref{eq:pv}) corresponds to calculations in two to two processes on the cross
section.
The results, properly normalized,
are process independent.}
\begin{eqnarray}
\frac{\partial \phi(z,t)}{\partial t}&=& \frac{g^2}{8\pi^2} \int^1_z \frac{dy}{y} \phi(z/y,t) 
P^V_{\phi \phi}(y)
\label{eq:app} 
\end{eqnarray}
The splitting functions are related by
$P_{\phi \phi}=P^R_{\phi \phi}+P^V_{\phi \phi}$, where
$R$ denotes the contribution from real boson emission.
$P_{\phi \phi}$ is free of logarithmic
corrections and positive definite.
The renormalizations with respect to the Born amplitude
as well as the ones belonging to the
next to leading terms at higher orders will be indicated below by
writing $\alpha(\overline{\mu}^2)
\longrightarrow \alpha(s)$.

Inserting the virtual probability of Eq. (\ref{eq:pv}) into the Eq.
(\ref{eq:app}) we find:
\begin{eqnarray}
\phi(1,t)&=& \phi_0 \exp \left\{ - \frac{ g^2(s)}{8 \pi^2} \left[ \left( T_i(T_i+1)+  \tan^2 \theta_{\rm w} 
\left( \frac{Y_i}{2} \right)^2 \right) \left( \log^2 \frac{s}{\mu^2}
 - 4 \log \frac{s}{\mu^2} \right) \right. \right. \nonumber \\
 && \left. \left. \;\;\;\;\;\;\;\;\;\;\;\;\;\;\;\;\;\;\;\;\;\;\;\; + 
 \frac{3}{2} \frac{m^2_t}{M^2} \log \frac{s}{\mu^2} \right] \right\} \label{eq:psol} 
\end{eqnarray}
\noindent These functions describe the total contribution for the emission of virtual particles (i.e. $z=1$),
with all invariants large compared to the cutoff $\mu$, to the
densities $\phi(z,t)$ ($\phi = \{ \phi^\pm, \chi, {\rm H} \}$). 
The normalization is not per line but on the level of the cross section.
For the invariant matrix element involving $n_\phi$ external 
scalar particles we thus find at the subleading level:
\begin{eqnarray}
&& {\cal M} (p_1,...,p_n,g_s,\mu^2)={\cal M} (p_1,...,p_n,g_s(s)) 
\exp \left\{ - \frac{1}{2}
\sum^{n_\phi}_{i=1} W^\phi_i(s,\mu^2) \right\}
\label{eq:mgyuk}
\end{eqnarray}
where
\begin{equation}
 W^\phi_i(s,\mu^2) =  \frac{ g^2(s)}{16 \pi^2} \!\! \left[ \! \left( T_i(T_i+1)+  \tan^2 \! 
 \theta_{\rm w}  
 \frac{Y^2_i}{4} \right) \!\! \left( \log^2 \frac{s}{\mu^2}- 4 \log \frac{s}{\mu^2} 
 \! \right) \!\!
 + \frac{3}{2} \frac{m^2_t}{M^2} \log \frac{s}{\mu^2} \right] \label{eq:Wphi}
\end{equation}
Again we note that the running coupling notation in the Born-amplitude of Eq. (\ref{eq:Wphi}) denotes
the renormalization corrections of the Born amplitude and higher order corrections should
be included by inserting a running coupling as in QCD \cite{ms2}.
The functions $W^\phi_i$ correspond to the probability of emitting a virtual
soft and/or collinear gauge
boson from the particle $\phi$ subject to the infrared cutoff $\mu$. Typical diagrams contributing
to Eq. (\ref{eq:Wphi}) in a covariant gauge are depicted in Fig. \ref{fig:nll}.
The universality of the splitting functions is crucial in obtaining the above result.

In addition to the Sudakov corrections in Eq. (\ref{eq:Wphi}) we also have to include terms
corresponding to the renormalization of the mass terms in the Yukawa coupling of the
Born amplitude $\left( \sim \frac{m_t^2}{M^2}, \frac{m_H^2}{M^2} \right)$ 
at the one loop level \cite{dp}. At
higher orders, mass renormalization terms are connected to two point functions and thus
subsubleading.

\section{Subleading corrections from Gribov's factorization theorem} \label{sec:gt}

In this section we present further evidence for the results of the previous section
by employing a non-Abelian version of Gribov's factorization theorem \cite{vg}. 
While in Ref. \cite{vg} only real bremsstrahlung corrections are discussed, the
form of the virtual soft and collinear divergences must factorize analogously due
to the KLN-theorem \cite{k,ln}. The non-Abelian
version is discussed in Ref. \cite{flmm}. The essential point is that to DL accuracy
the soft and collinear
gauge boson with the smallest $|{\mbox{\boldmath $k$}_\perp}|$ factorizes. From the definition
in Eq. (\ref{eq:kpdef}) it is clear that ${\mbox{\boldmath $k$}}^2_\perp \approx |k|^2 \theta^2$, 
where $\theta$ denotes the angle between the emitted gauge boson of momentum $k$
and the external line emitting this boson, i.e. both
soft as well as collinear emission contributions are regularized simultaneously.
In the Feynman gauge we then have \cite{flmm}:
\begin{eqnarray}
{\cal M}(p_1,...,p_n;\mu^2) & = & {\cal M}_{\rm Born}(p_1,...,p_n) -\frac{i}{2}
\frac{g^2_s}{(2\pi)^4} \sum_{j,l=1, j \neq l}^n \int_{s \gg \mbox{{\scriptsize \boldmath $k$}}^2_\perp
\gg \mu^2} \frac{d^4k}{k^2+i \epsilon} \;\;
\frac{p_jp_l}{(kp_j)(kp_l)}  \nonumber \\
& & \times \; T^a(j) T^a(l) {\cal M} (p_1,...,p_n;{\mbox{\boldmath $k$}%
^2_\perp}) \,,  \label{eq:vem}
\end{eqnarray}
From Eq. (\ref{eq:kpdef}) it is clear that $\frac{p_jp_l}{(kp_j)(kp_l)}= 
\frac{2}{{\mbox{\boldmath $k$}^2_\perp}}$ 
and that Eq. (\ref{eq:vem}) has the required factorized form.
For the DL corrections
it is convenient to employ the Sudakov parametrization \cite{s} given by:
\begin{equation}
k=vp_j \; + u p_l + k_\perp \label{eq:sudp}
\end{equation}
For the boson propagator we use the identity
\begin{equation}
\frac{i}{k^2+i\varepsilon}=
\frac{i}{s u v -{\mbox{\boldmath $k$}^2_{\perp}}+i\varepsilon}= {\cal P}
\frac{i}{s u v -{\mbox{\boldmath $k$}^2_{\perp}}} + \pi \delta ( s u v -
{\mbox{\boldmath $k$}^2_{
\perp}}) \label{eq:propid}
\end{equation}
writing it in form of the real and imaginary parts (the principle value is indicated
by ${\cal P}$).
The latter does not contribute to the DL asymptotics and at higher orders gives
subsubleading
contributions. The cutoff will be introduced via the function $\Theta ( 
{\mbox{\boldmath $k$}^2_{\perp}} - \mu^2)$.
Rewriting the measure as $d^4k=d^2k_\perp d^2k_\parallel$ with
\begin{eqnarray}
d^2k_\perp &=& |{\mbox{\boldmath $k$}_{\perp}}| d |{\mbox{\boldmath $k$}_{\perp
}}| d \varphi =
\frac{1}{2} d {\mbox{\boldmath $k$}^2_{\perp}} d \varphi = \pi d {\mbox{\boldmath
$k$}^2_{\perp}} \\
d^2k_\parallel &=& | \partial (k^0,k^x)/ \partial (u,v)| d u d v 
= | p_{j_0}p_{l_x}-p_{l_0}p_{j_x}| du dv \approx \frac{
s}{2} du dv
\end{eqnarray}
where we turn the coordinate system such that the $p_j,p_l$ plane corresponds to
$0,x$ and the
$y,z$ coordinates to the $k_\perp$ direction so that it is purely spacelike.
The last equation follows from $p_i^2=0$, i.e. $p_{i_x}^2\approx p_{i_0}^2$ and
\begin{equation}
(p_{j_0}p_{l_x}-p_{l_0}p_{j_x})^2 \approx (p_{j_0}p_{l_0}-p_{l_x}p_{j_x})^2=(p_j
p_l)^2=(s/2)^2
\end{equation}
Using in addition the
conservation of the total group charge 
\begin{equation}
\sum_{j=1}^n T^a(j) {\cal M}(p_1,...,p_j,...,p_n;{\mbox{\boldmath
$k$}_{\perp }^{2}})=0
\end{equation}
we arrive at:
\begin{eqnarray}
{\cal M}(p_{1},...,p_{n};\mu ^{2}) &=&{\cal M}_{\rm Born}(p_{1},...,p_{n})-\frac{%
2g^{2}_s}{(4\pi )^{2}}\sum_{l=1}^{n}\int_{\mu ^{2}}^{s}\frac{d{%
\mbox{\boldmath $k$}_{\perp }^{2}}}{{\mbox{\boldmath $k$}_{\perp
}^{2}}}\int_{|\mbox{\scriptsize \boldmath $k$}_{\perp }|/\sqrt{s}}^1
\frac{dv}{v}  \nonumber \\
&&\times \;C_{l}{\cal M}(p_{1},...,p_{n};{\mbox{\boldmath
$k$}_{\perp }^{2}}%
)\;\;,  \label{eq:dgvem}
\end{eqnarray}
where $C_{l}$ is the eigenvalue of the Casimir operator $T^{a}(l)T^{a}(l)$
(In QCD $%
C_{l}=C_{A}$ for gauge bosons in the adjoint representation of the gauge
group $SU(N)$ and $C_{l}=C_{F}$ for fermions in the fundamental
representation).
The differential form of the infrared evolution equation follows immediately
from (\ref{eq:dgvem}):
\begin{equation}
\frac{\partial {\cal M}(p_{1},...,p_{n};\mu ^{2})}{\partial \log (\mu
^{2})}%
=K(\mu ^{2}){\cal M}(p_{1},...,p_{n};\mu ^{2})\,,  \label{eq:ee}
\end{equation}
where
\begin{equation}
K(\mu ^{2})\equiv -\frac{1}{2}\sum_{l=1}^{n}\frac{\partial W_{l}(s,\mu
^{2})%
}{\partial \log (\mu ^{2})}
\end{equation}
with
\begin{equation}
W_{l}(s,\mu ^{2})=\frac{g^{2}_s}{(4\pi)^{2}}C_{l}\, \log^2 \frac{s}{\mu^2}
\,.  \label{eq:wl}
\end{equation}
$W_l$ is the probability to emit a soft and
almost collinear gauge boson from the particle $l$,
subject to the infrared cut-off $\mu $ on the transverse momentum \cite{flmm}. Note
that in distinction to a gluon or photon mass regulator, the cut-off $\mu$ 
does not spoil the gauge invariance of the theory
and can take on arbitrary values, i.e. it is not necessarily taken to zero.
To logarithmic accuracy, we obtain directly from (\ref{eq:wl}):
\begin{equation}
\frac{\partial W_{l}(s,\mu ^{2})}{\partial \log (\mu ^{2})}=-\frac{g^{2}_s}{%
8\pi ^{2}}C_{l}\log \frac{s}{\mu^2}\,.
\end{equation}
The infrared evolution equation (\ref{eq:ee}) should be solved with an
appropriate initial condition. In the case of large scattering angles, if we
choose the cut-off to be the large scale $\sqrt{s}$ then clearly there are no
Sudakov corrections. The initial condition is therefore
\begin{equation}
{\cal M}(p_{1},...,p_{n};s)={\cal M}_{\rm Born}(p_{1},...,p_{n}),
\end{equation}
and the solution of (\ref{eq:ee}) is thus given by the product of the Born
amplitude and the Sudakov form factors:
\begin{equation}
{\cal M}(p_{1},...,p_{n};\mu ^{2})={\cal M}_{\rm Born}(p_{1},...,p_{n})\exp
\left( -\frac{1}{2}\sum_{l=1}^{n}W_{l}(s,\mu ^{2})\right) \label{eq:NAexp}
\end{equation}
This is the exponentiation of Sudakov DL in non-Abelian gauge theories \cite{NA}.
We now want to apply this result to the electroweak theory for the subleading Yukawa
corrections at higher orders.
\begin{figure}
\centering
\epsfig{file=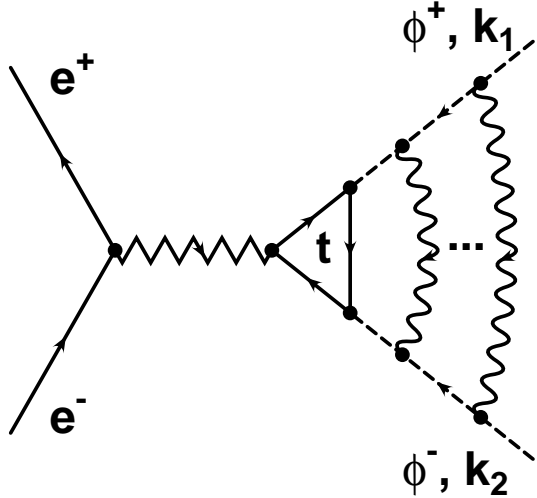,width=7cm}
\caption{A Feynman diagram yielding Yukawa enhanced logarithmic corrections in the
on-shell scheme. At higher orders, the subleading corrections are given in factorized
form according to the non-Abelian generalization of Gribov's theorem as described in the
text. Corrections from gauge bosons inside the top-loop give only sub-sub leading contributions.
DL-corrections at two and higher loop order are given by gauge bosons coupling to (in principle
all) external legs as schematically indicated.}
\label{fig:yuk}
\end{figure}
Since we are interested here in corrections to order ${\cal O}(\alpha^n L^{2n-1})$, each additional
loop correction to the universal subleading terms in the previous section must yield two logarithms,
i.e. we are considering DL-corrections to the basic process like the inner fermion loop
in Fig. \ref{fig:yuk}.
It is of particular importance that all additional gauge bosons must couple to external legs, since
otherwise only a subleading term of order ${\cal O}(\alpha^n L^{2n-2})$ would be generated. 
All subleading corrections generated by the exchange of gauge bosons coupling both to
external Goldstone bosons and inner fermion lines cancel analogously to a mechanism found
in Ref. \cite{ms1} for terms in heavy quark production in $\gamma \gamma$-collisions in a
$J_z=0$ state. Formally this can be understood by noting that such terms contain an infrared
divergent correction. The sum of those terms, however, is given by the Sudakov form factor.
Thus any additional terms encountered in intermediate steps of the calculation cancel.
For the one loop process in Fig. \ref{fig:yuk}, for instance,
we inner include only corrections with top quarks and assume
on-shell renormalization of the external Goldstone bosons. 
Thus the corrections at higher orders factorize with respect to the one loop fermion amplitude
and ${\cal M}_{\rm ``Born''}(p_1,...,p_n) = {\cal M}_{\rm 1 loop}(p_1,...,p_n)$. Note that
the latter is also independent of the cutoff $\mu$ since the fermion mass serves as a
natural regulator. In principle we can choose the top-quark mass to be much larger than $\mu$
for instance. This freedom is not present for subleading terms from gauge bosons, such
as the angular contributions of the type $\log \frac{u}{t} \log \frac{s}{M^2}$, which furthermore
don't factorize with respect to the Born amplitude as mentioned above. 
Thus, the method suggested in Ref. \cite{kps}
for the higher order angular terms  
cannot straightforwardly be justified via the non-Abelian
generalization of Gribov's bremsstrahlung theorem\footnote{Using the
${\overline {\rm MS}}$-renormalization scheme, however, the subleading pole structure of QCD scattering
amplitudes at the two loop level is determined only by one loop divergences and renormalization
group corrections \cite{cat}.}.
In our case we have for the two loop electroweak DL corrections at the weak scale $\mu=M$: 
\begin{equation}
W^{\rm ew}_l (s , M^2) = 
  \frac{ g^2(s)}{16 \pi^2} \!\! \left[ \! \left( T_i(T_i+1)+  \tan^2 \! 
 \theta_{\rm w}  
 \frac{Y^2_i}{4} \right)  \log^2 \frac{s}{M^2}  
 \right] \label{eq:Wew}
\end{equation}
We now want to consider specific processes relevant at future $e^+e^-$ colliders and
demonstrate how to apply the non-Abelian version of Gribov's factorization theorem for
the higher order corrections. The subleading corrections are then compared to the 
general splitting
function approach of section \ref{sec:sf}.

In the case of the amplitude of Fig. \ref{fig:yuk} we must use the quantum numbers of the
associated Goldstone bosons and we have the following Born amplitude
\begin{equation}
{\cal M}_{\rm Born}(p_{1},...,p_{4})=i \frac{e^2}{2s c^2_{\rm w}} \langle e^-_{\rm R}|
\gamma^\nu | e^+_{\rm L} \rangle (k_1-k_2)_\nu 
\end{equation}
and at one loop we have two fermion loops contributing ($ttb$ and $bbt$). 
The renormalization condition is provided by the requirement that the corrections
vanish at the weak scale, i.e. for $s=M^2$, which amounts to subtracting the
vertex for that case. The first diagram of the two is given by
\begin{eqnarray}
&&{\cal A}^{ttb}_{\rm 1 loop}(p_{1},...,p_{4})= 3 \sum_{\gamma, Z} \frac{e^4 m_t^2}{2 s M^2 s^2_{\rm w}} 
\langle e^-_{\rm R}| \gamma^\nu | e^+_{\rm L} \rangle c_+^e \times \nonumber \\
&& \int \frac{d^nl}{(2 \pi)^n} \frac{ {\rm Tr} \left\{ \omega_- {\rlap/ l} \omega_+ (
{\rlap/ l}-{\rlap/ k_2}) \gamma_\nu (c^t_+ \omega_+ + c^t_- \omega_-) ( {\rlap/ l}+ {\rlap/ k_1}) 
\right\}}{(l^2-m_b^2+i \varepsilon)((l+k_1)^2-m_t^2+i \varepsilon)((l-k_2)^2-m_t^2+i \varepsilon)}
+ \delta^{ttb}_{\rm ct}
\nonumber \\ &&= \frac{3iQ_t}{16 \pi^2 c^2_{\rm w}} \frac{e^4 m_t^2}{2 s M^2 s^2_{\rm w}}
\langle e^-_{\rm R}| \gamma^\nu | e^+_{\rm L} \rangle (B_{23}-B_{23}^M)
(k_1-k_2)_\nu \label{eq:ttb}
\end{eqnarray}
where $\omega_\pm=\frac{1}{2} ( 1 \pm \gamma_5)$ and the chiral couplings are given by
$c^f_\pm=Q_f$ for the photon and $c^f_+=\frac{s_{\rm w}}{c_{\rm w}}Q_f$ and
$c^f_-= \frac{s_{\rm w}^2 Q_f-T^3_f}{s_{\rm w}c_{\rm w}}$ for Z-bosons respectively.
The counterterm $\delta^{ttb}_{\rm ct}$ 
is chosen such that the logarithmic corrections vanish for $s=M^2$.
Thus, the sum of the scalar functions is to logarithmic accuracy $B_{23}-B_{23}^M=-\log \frac{s}{M^2}$.
Analogously, we have for the $bbt$ quark loop:
\begin{eqnarray}
&&{\cal A}^{bbt}_{\rm 1 loop}(p_{1},...,p_{4})= 3 \sum_{\gamma, Z} \frac{e^4 m_t^2}{2 s M^2 s^2_{\rm w}} 
\langle e^-_{\rm R}| \gamma^\nu | e^+_{\rm L} \rangle c_+^e \times \nonumber \\
&& \int \frac{d^nl}{(2 \pi)^n} \frac{ {\rm Tr} \left\{ \omega_+ (-{\rlap/ l}) \omega_- (-
{\rlap/ l}-{\rlap/ k_1}) \gamma_\nu (c^b_+ \omega_+ + c^b_- \omega_-) ( -{\rlap/ l}+ {\rlap/ k_2}) 
\right\}}{(l^2-m_t^2+i \varepsilon)((l+k_1)^2-m_b^2+i \varepsilon)((l-k_2)^2-m_b^2+i \varepsilon)}
+ \delta^{bbt}_{\rm ct}
\nonumber \\ &&= -\frac{3i(Q_b-T^3_b)}{16 \pi^2 c^2_{\rm w}} \frac{e^4 m_t^2}{2 s M^2 s^2_{\rm w}}
\langle e^-_{\rm R}| \gamma^\nu | e^+_{\rm L} \rangle (B_{23}-B_{23}^M)
(k_1-k_2)_\nu \label{eq:bbt}
\end{eqnarray}
Adding both results (\ref{eq:ttb}) and (\ref{eq:bbt}) we find
\begin{equation}
{\cal M}_{\rm 1 loop}(p_{1},...,p_{4}) = 
{\cal M}_{\rm Born}(p_{1},...,p_{4}) \left\{ 1 - 
\frac{g^2}{16 \pi^2} \frac{3}{2} \frac{m_t^2}{M^2} \log \frac{s}{M^2} \right\} \label{eq:yuk}
\end{equation}
and the all orders result to subleading accuracy is given by
\begin{equation}
{\cal M}(p_{1},...,p_{n};\mu ^{2})={\cal M}_{\rm 1 loop}(p_{1},...,p_{n})\exp
\left( -\frac{1}{2}\sum_{l=1}^{n}W^{\rm ew}_{l}(s,\mu ^{2})\right) \label{eq:yukg}
\end{equation}
The subleading Yukawa corrections from the Altarelli-Parisi in Eq. (\ref{eq:mgyuk}) agree with
the corresponding results from the application of the Gribov-theorem in Eq. (\ref{eq:yukg}).
\begin{figure}
\centering
\epsfig{file=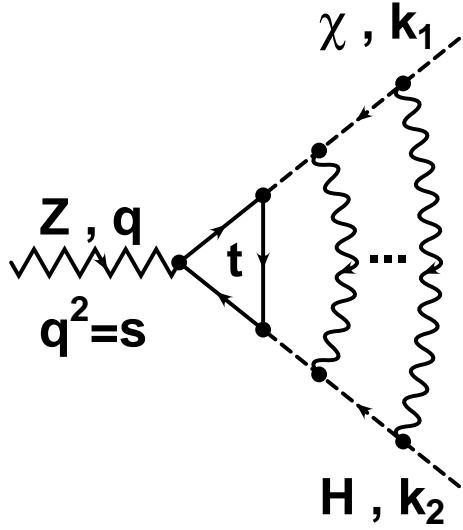,width=6cm}
\caption{A Feynman diagram yielding Yukawa enhanced logarithmic corrections to external
longitudinal Z-bosons and Higgs lines in the
on-shell scheme. At higher orders, the subleading corrections are given in factorized
form according to the non-Abelian generalization of Gribov's theorem as described in the
text. Corrections from gauge bosons inside the top-loop give only sub-sub leading contributions.}
\label{fig:ZchiH}
\end{figure}
For longitudinal Z-boson and Higgs production, we note that there is only one non-mass suppressed
elementary vertex with two neutral scalars, 
namely the $Z \chi H$ vertex. As mentioned above, universal terms are related
to the massless limit. For the ``Born amplitude'' of the Higgs-Strahlung vertex we have
\begin{equation}
{\cal M}^{\rm Z \chi H}_{\rm Born} = \frac{e}{2 s_{\rm w} c_{\rm w}} (k_1^\nu - k_2^\nu)
\end{equation}
The universal Yukawa corrections to both external $\chi$ and $H$ states from an off shell
$Z$ line are then given by the corrections depicted in the inner fermion
loop of Fig. \ref{fig:ZchiH}. Here we find
\begin{eqnarray}
&&{\cal A}^{Z \chi H}_{\rm 1 loop}(p_{1},...,p_{3})= 3 \frac{e^3 m_t^2}{4 M^2 s^2_{\rm w}} 
\times \nonumber \\
&& \int \frac{d^nl}{(2 \pi)^n} \frac{ {\rm Tr} \left\{ \gamma_5 ({\rlap/ l}) (
{\rlap/ l}-{\rlap/ k_2}) \gamma^\nu (c^t_+ \omega_+ + c^t_- \omega_-) ( {\rlap/ l}+ {\rlap/ k_1}) 
\right\}}{(l^2-m_t^2+i \varepsilon)((l+k_1)^2-m_t^2+i \varepsilon)((l-k_2)^2-m_t^2+i \varepsilon)}
+ \delta^{Z \chi H}_{\rm ct}
\nonumber \\ &&= \frac{6T^3_t}{16 \pi^2 s_{\rm w} c_{\rm w}} \frac{e^3 m_t^2}{4 M^2 s^2_{\rm w}}
(B_{23}-B_{23}^M)
(k^\nu_1-k^\nu_2) 
\end{eqnarray}
and thus
\begin{equation}
{\cal M}^{Z \chi H}_{\rm 1 loop}(p_{1},...,p_{3}) = {\cal M}^{\rm Z \chi H}_{\rm Born} \left\{ 1
- \frac{3}{2} \frac{e^2 m^2_t}{ 16 \pi^2 s^2_{\rm w} M^2} 
\log \frac{s}{M^2} \right\} \label{eq:ZchiH}
\end{equation}
From the same line of reasoning as for the charged Goldstone bosons we find that the
all orders result is given by Eq. (\ref{eq:yukg}). At the subleading level, this is equivalent to
the corresponding corrections obtained in Eq. (\ref{eq:mgyuk}).

\section{Top Yukawa corrections for chiral quark production} \label{sec:cf}

In Ref. \cite{m1} all subleading Sudakov logarithms were resummed assuming that all invariants
$2p_jp_l \sim s$. The subleading kernel of the infrared evolution equation was determined by
using the virtual contributions to the splitting functions from QCD and applying these results
to the high energy regime of the electroweak theory. Soft photon corrections were then added
by appropriate matching conditions at the weak scale. We explicitly restricted ourselves in
Ref. \cite{m1} to the case where all fermions had masses below the weak scale and thus excluded
Yukawa enhanced terms. From the arguments of section \ref{sec:gt} it is now straightforward
to include also top-Yukawa terms for chiral quark final states. These terms occur for
left handed bottom as well as top quark external lines. The situation for a typical Drell-Yan
process is depicted in Fig. \ref{fig:tyuk} where for the inner scattering amplitude we have
two contributions. We neglect all terms of order ${\cal O} \left( \frac{m_f^2}{s}, \frac{M^2}{s}
\right)$. Using on-shell renormalization we find for the inner amplitude on the left
in Fig. \ref{fig:tyuk} for a right handed electron in the initial and a left handed bottom quark
in the final state from the $\phi^\pm$ loop for the sum of the $\gamma$ and $Z$ contributions:
\begin{figure}
\centering
\epsfig{file=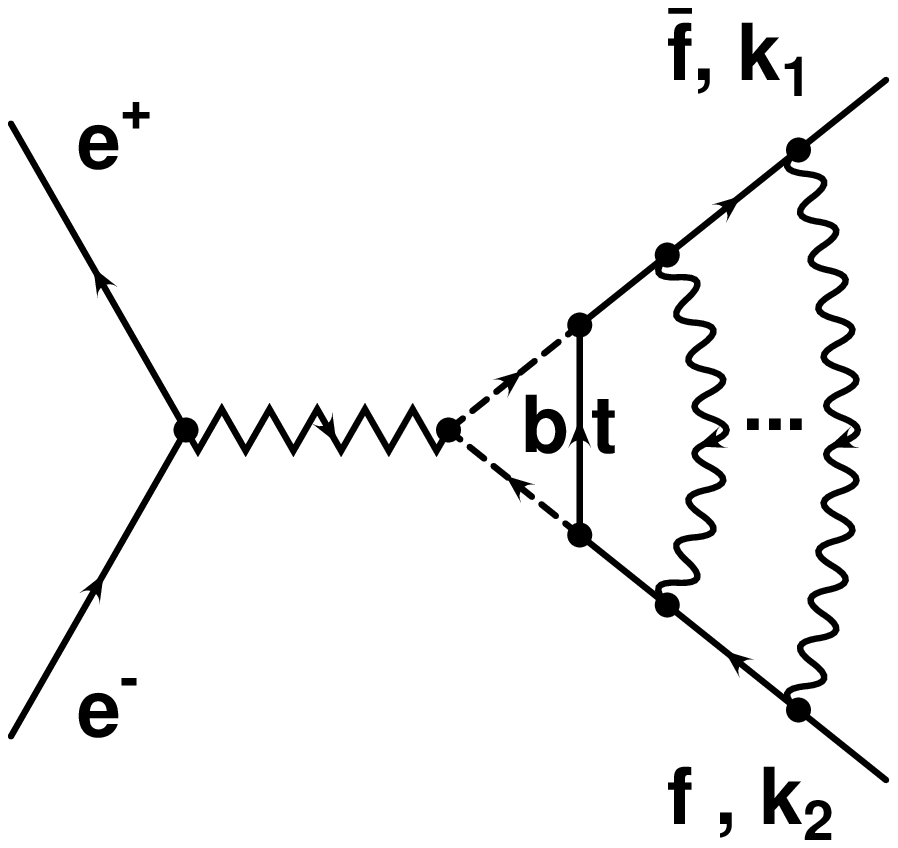,width=6cm} \hspace{2cm}
\epsfig{file=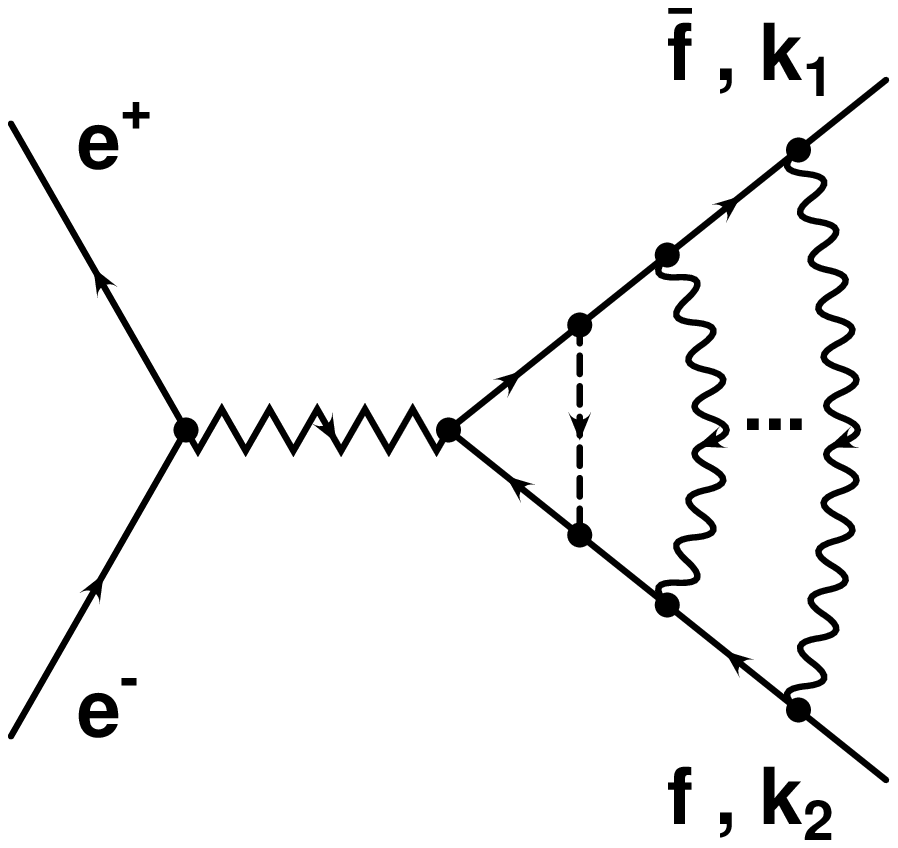,width=6cm}
\caption{Feynman diagrams yielding Yukawa enhanced logarithmic corrections to the third
generation of fermions in the final sate. The inner scattering amplitude 
is taken on the mass shell. No DL-corrections originate from the inner loop.
At higher orders, the subleading corrections are given in factorized
form according to the non-Abelian generalization of Gribov's theorem as described in the
text. Corrections from gauge bosons inside the Goldstone-boson loop give only 
sub-sub leading contributions.
DL-corrections at two and higher loop order are given by gauge bosons coupling to (in principle
all) external legs as schematically indicated.}
\label{fig:tyuk}
\end{figure}
\begin{eqnarray}
{}^a\!\!{\cal A}^{\rm DY}_{\rm 1 loop} &=& - \frac{e^4m_t^2}{4 s M^2 s_{\rm w}^2 c_{\rm w}^2} 
\langle e^+_{\rm L} | \gamma_\nu | e^-_{\rm R} \rangle \times \nonumber \\ &&
\int \frac{ d^nl}{(2 \pi)^n}
\frac{ \langle f_{\rm L} | {\rlap/ l} (2l-k_1-k_2)^\nu | f_{\rm R} \rangle}{(l^2-m_{f^\prime}^2+i \varepsilon
)((l-k_1)^2-M^2+i \varepsilon)((l-k_2)^2-M^2+i \varepsilon)}
+{}^a \delta^{\rm DY}_{\rm ct} \nonumber \\ &=&
-\frac{i}{32 \pi^2} \frac{e^4m_t^2}{4 s M^2 s_{\rm w}^2 c_{\rm w}^2}
\langle e^+_{\rm L} | \gamma_\nu | e^-_{\rm R} \rangle  \langle f_{\rm L} |
\gamma^\nu | f_{\rm R} \rangle (B_{23}-B^M_{23}) \label{eq:sfs}
\end{eqnarray}
The scalar functions at high energy 
evaluate to $B_{23}-B^M_{23}=-\log \frac{s}{M^2}$ as mentioned above. For the diagram on the
right in Fig. \ref{fig:tyuk} we have for the bottom again only the $\phi^\pm$ contribution.
Here we find for the sum of the $\gamma$ and $Z$ contributions:
\begin{eqnarray}
{}^b\!\!{\cal A}^{\rm DY}_{\rm 1 loop} &=& - \frac{e^4m_t^2Q_t}{2 s M^2 s_{\rm w}^2 c_{\rm w}^2} 
\langle e^+_{\rm L} | \gamma_\nu | e^-_{\rm R} \rangle \times \nonumber \\ &&
\int \frac{ d^nl}{(2 \pi)^n}
\frac{ \langle f_{\rm L} | {\rlap/ l} \gamma^\nu {\rlap/ l}| f_{\rm R} \rangle}{(l^2-M^2+i \varepsilon
)((l-k_1)^2-m_t^2+i \varepsilon)((l-k_2)^2-m_t^2+i \varepsilon)}
+{}^b \delta^{\rm DY}_{\rm ct} \nonumber \\ &=&
\frac{i}{32 \pi^2} \frac{e^4m_t^2Q_t}{2 s M^2 s_{\rm w}^2 c_{\rm w}^2}
\langle e^+_{\rm L} | \gamma_\nu | e^-_{\rm R} \rangle  \langle f_{\rm L} |
\gamma^\nu | f_{\rm R} \rangle (B_{23}-B^M_{23}) \label{eq:fsf}
\end{eqnarray}
In all cases we renormalize on-shell, i.e. by requiring that the vertex vanishes when the momentum
transfer equals the masses of the external on-shell lines. All on-shell self energy contributions
don't contribute in this scheme.
For external left handed top quarks, the $\phi^\pm$ loop is mass suppressed and we only have
to consider the $\chi$ and $H$ corrections. 
They are given by replacing $Q_t \longrightarrow 2 Q_t \left(T^3_t \right)^2$ and
$Q_t \longrightarrow \frac{1}{2} Q_t$ in Eq. (\ref{eq:fsf}). It turns out that the $Z \chi H$
contributions equal the corrections from the $\gamma \phi^\pm$ and $Z \phi^\pm$ in the case
of the bottom calculation.
The Born amplitude is given by:
\begin{eqnarray}
{\cal M}^{\rm DY}_{\rm Born} &=&
i \frac{e^2}{s c^2_{\rm w}} (Q_f - T^3_f) \langle e^+_{\rm L} | \gamma_\nu | e^-_{\rm R} \rangle  
\langle f_{\rm L} | \gamma^\nu | f_{\rm R} \rangle \nonumber \\
&=& \left\{ \begin{array}{lc}
i \frac{e^2}{6 s c^2_{\rm w}} \langle e^+_{\rm L} | \gamma_\nu | e^-_{\rm R} \rangle  
\langle f_{\rm L} | \gamma^\nu | f_{\rm R} \rangle  \;\;\;, f_{\rm L}=t_{\rm L}, b_{\rm L} \\
i \frac{e^2}{s c^2_{\rm w}} \frac{2}{3} \langle e^+_{\rm L} | \gamma_\nu | e^-_{\rm R} \rangle  
\langle f_{\rm R} | \gamma^\nu | f_{\rm L} \rangle \;\;, f_{\rm R}=t_{\rm R} \end{array} \right.
\label{eq:LB}
\end{eqnarray}
for top and bottom quarks.
In all cases, $\log \frac{M^2}{m_t^2}$
terms can be savely neglected to the accuracy we are working.
Thus we find for left handed quarks of the third generation:
\begin{eqnarray}
{\cal M}^{{\rm DY}_{\rm L}}_{\rm 1 loop}(p_{1},...,p_{4})&=&{\cal M}^{\rm DY}_{\rm Born}(p_{1},...,p_{4}) \left\{ 1 - 
\frac{g^2}{16 \pi^2} \frac{1}{4} 
\frac{m_t^2}{M^2} \delta_{f,t_{\rm L}/b_{\rm L}} \log \frac{s}{M^2} 
\right\} \label{eq:tyukL}
\end{eqnarray}
For right handed external top quarks we have $\phi^\pm$, $\chi$ and $H$ corrections. In that case we
observe that the $Z \chi H$, $\gamma \phi^\pm$ and $Z \phi^\pm$ loops have an opposite sign
relative to the left handed case.
For the corrections corresponding to the topology shown on the right in Fig. \ref{fig:tyuk}
we must replace $Q_t$ in Eq. (\ref{eq:fsf}) by $Q_f-T^3_f=\frac{1}{6}$ for the $\phi^\pm$ graph.
The same contribution is obtained by adding the $H$ and $\chi$ loops and we find:
\begin{equation}
{\cal M}^{{\rm DY}_{\rm R}}_{\rm 1 loop}(p_{1},...,p_{4})={\cal M}^{\rm DY}_{\rm Born}(p_{1},...,p_{4}) \left\{ 1 - 
\frac{g^2}{16 \pi^2} \frac{1}{2} 
\frac{m_t^2}{M^2} \delta_{f,t_{\rm R}} \log \frac{s}{M^2} 
\right\} \label{eq:tyukR}
\end{equation}
At higher orders
we note that the exchange of gauge bosons inside the one loop process is subsubleading
and we arrive at the factorized form analogous to the Yukawa corrections in section \ref{sec:gt}.
Since these corrections are of universal nature we can drop the specific reference to the
Drell-Yan process and
the application of the generalized Gribov-theorem for external fermion lines
to all orders yields:
\begin{equation}
{\cal M} (p_{1},...,p_{n};\mu ^{2})={\cal M}_{\rm 1 loop}(p_{1},...,p_{n})\exp
\left( -\frac{1}{2}\sum_{l=1}^{n_f}W^{\rm ew}_{l}(s,\mu ^{2})\right) \label{eq:gt}
\end{equation}
where $W^{\rm ew}_{l}(s,\mu ^{2})$ is given in Eq. (\ref{eq:Wew}) and the quantum numbers are
those of the external fermion lines.
Since at high energies all fermions can be considered massless we can again absorb the chiral
top-Yukawa corrections into universal splitting functions as in Ref. \cite{m1}. Thus in the electroweak
theory we find to next to leading order the corresponding probability for the emission of
gauge bosons from chiral fermions subject to the cutoff $\mu$:
\begin{eqnarray}
 W^f_i(s,\mu^2) &=&  \frac{ g^2(s)}{16 \pi^2} \!\! \left[ \! \left( T_i(T_i+1)+  \tan^2 \! 
 \theta_{\rm w}  
 \frac{Y^2_i}{4} \right) \!\! 
\left( \log^2 \frac{s}{\mu^2}- 3 \log \frac{s}{\mu^2} 
 \! \right) \right. \nonumber \\ && \left.
 + \left( \frac{1+\delta_{f,{\rm R}}}{4} \frac{m^2_f}{M^2} + \delta_{f,{\rm L}} 
 \frac{m^2_{f^\prime}}{4 M^2} \right) 
 \log \frac{s}{\mu^2} \right] \label{eq:Wf}
\end{eqnarray}
The second line only contributes for left handed bottom and for top quarks as mentioned above and
$f^\prime$ denotes the corresponding isospin partner for left handed fermions.

\section{Semi-inclusive cross sections} \label{sec:si}

Up to this point we have only considered the corrections from virtual corrections above the
weak scale $M$. The physical photon, however, is massless and must be included in a semi-inclusive
or fully inclusive way. It is thus necessary to consider now the regime for 
${\mbox{\boldmath $k$}^2_{\perp}} < M^2$. The corrections for external fermion, photon and
$W^\pm$ lines are given in Ref. \cite{m1}, in each case corresponding to the logarithmic 
probability to emit soft and/or collinear particles below the scale $M$. The high energy
solution is then the boundary condition for the infrared evolution equation at the scale
$\mu=M$. For the longitudinal particles, we only have corrections from the charged gauge
bosons below the scale $M$. In this regime we also need to consider particle masses.
For real photon emission we assume that the detector resolution is bounded by $\mu_{\rm expt}
<M$, so that emission from real massive gauge bosons does not need to be considered and
for simplicity, we restrict ourselves here to the soft photon approximation. 

Under these circumstances we are now able to summarize the complete expression for observable
electroweak cross sections at high energies for all universal leading and subleading Sudakov
corrections as follows\footnote{We emphasize that for photon and Z-boson final states
the mixing effects have to be included correctly as described in Ref. \cite{m1}. In particular,
for transverse degrees of freedom the corrections don't factorize with respect to the 
physical Born amplitude
but rather with respect to the amplitudes containing the fields in the broken phase. 
For longitudinally polarized Z-bosons, however, there is no mixing with photons
and the corrections factorize with respect to the Born amplitude.}:
\begin{eqnarray}
&& d\sigma (p_{1}, \ldots, p_{n},g,g^\prime,\mu_{exp}) = d\sigma_{\rm Born} (p_{1},
\ldots ,p_{n},g(s),g^\prime (s))
\nonumber \\ && \times \exp \left\{ - \sum^{n_g}_{i=1} W^g_i (s,M^2)
- \sum^{n_f}_{i=1} W^f_i (s,M^2) - \sum^{n_\phi}_{i=1} W^\phi_i (s,M^2)
\right\} \nonumber \\
&&\times \exp \left[ - \sum_{i=1}^{n_f} \left( w^f_i(s,\mu^2)
- w^f_i(s,M^2) \right)
- \sum_{i=1}^{n_w} \left( w^{\rm w}_i(s,\mu^2)
- w^{\rm w}_i(s,M^2) \right) \right. \nonumber \\
&& \;\;\;\;\;\;\;\;\;\;\; \left. - \sum_{i=1}^{n_\gamma} w_i^\gamma(M^2,m_j^2)
\right] 
\times \exp \left( w^\gamma_{\rm expt} (s,m_i,\mu,\mu_{\rm expt})
\right) \label{eq:si}
\end{eqnarray}
The functions $W^\phi_i (s,M^2)$ and $ W^f_i (s,M^2)$ are given in Eqs. (\ref{eq:Wphi}) and
(\ref{eq:Wf}) respectively. The remaining logarithmic probabilities are given in Ref.
\cite{m1} and are summarized for convenience below:
\begin{eqnarray}
W^g_i(s,M^2) &=& \left( \frac{\alpha(s)}{4 \pi}T_i(T_i+1)+ \frac{\alpha^\prime(s)}{4 \pi}
\left( \frac{Y_i}{2} \right)^2 \right) \log^2 \frac{s}{M^2} \nonumber \\
&& 
- \left( \delta_{i,{\rm W}} \frac{\alpha(s)}{\pi} \beta_0 + \delta_{i,{\rm B}}
\frac{\alpha^\prime(s)}{\pi} \beta^\prime_0 \right) \log \frac{s}{M^2} \label{eq:Wg}
\end{eqnarray}
with
\begin{equation}
\beta_0=\frac{11}{12}C_A - \frac{1}{3}n_{gen}-\frac{1}{24}n_{h} \;\;\;,\;\;\;
\beta^\prime_0= - \frac{5}{9}n_{gen} -\frac{1}{24}n_{h} \label{eq:ewb0}
\end{equation}
where $n_{gen}$ denotes the number of fermion generations \cite{wein,gross} and $n_h$ the number of
Higgs doublets. Again we note that for external photon and Z-boson states we must include
the mixing appropriately as discussed in Ref. \cite{m1}.
For the terms entering from contributions below the weak scale we have for fermions:
\begin{equation}
w^f_i(s,\mu^2) = \left\{ \begin{array}{lc} \frac{e_i^2}{(4 \pi)^2} \left( \log^2 \frac{s}{\mu^2}
- 3 \log \frac{s}{\mu^2} \right) & , \;\;\; m_i \ll \mu \\
\frac{e_i^2}{(4 \pi)^2} \left[ \left( \log \frac{s}{m_i^2}-1 \right) 2 \log \frac{m_i^2}{\mu^2} 
\right. \\
\left.\;\;\;\;\;\;\;\;\;\;+ \log^2 \frac{s}{m_i^2} - 3 \log \frac{s}{m_i^2} \right] & , \;\;\; \mu 
\ll m_i\end{array} \right.
\end{equation}
Analogously, for external W-bosons and photons we find:
\begin{equation}
w^{\rm w}_i(s,\mu^2) =
\frac{e_i^2}{(4 \pi)^2} \left[ \left( \log \frac{s}{M^2}-1 \right)
2 \log \frac{M^2}{\mu^2}
+ \log^2 \frac{s}{M^2} \right]
\end{equation}
\begin{equation}
w_i^\gamma(M^2,\mu^2) = \left\{ \begin{array}{lc}
\frac{1}{3} \sum_{j=1}^{n_f} \frac{e_j^2}{4 \pi^2} N^j_C
\log \frac{M^2}{\mu^2} & , \;\;\; m_j \ll \mu \\
\frac{1}{3} \sum_{j=1}^{n_f} \frac{e_j^2}{4 \pi^2} N^j_C \log \frac{M^2}{m_j^2}
& , \;\;\; \mu \ll m_j\end{array} \right.
\end{equation}
for the virtual corrections and for real photon emission we have in the soft
photon approximation:
\begin{eqnarray}
w_{\rm expt}^{\gamma }(s,m_i,\mu,\mu_{\rm expt})
\!\!&=&\!\!\! \left\{ \begin{array}{lc}
\sum_{i=1}^n \frac{e_i^2}{(4 \pi)^2} \left[
- \log^2 \frac{s}{\mu^2_{\rm expt}}
+ \log^2 \frac{s}{\mu^2}- 3 \log \frac{s}{\mu^2} \right]
& , m_i \ll \mu \\
\sum_{i=1}^n \frac{e_i^2}{(4 \pi)^2} \left[ \left( \log
\frac{s}{m_i^2} -  1 \right)
2 \log \frac{m_i^2}{\mu^2} + \log^2 \frac{s}{m_i^2}
\right. \\ \left. - 2 \log \frac{s}{\mu^2_{\rm expt}} \left( \log \frac{s}{m_i^2} -
1 \right) \right]
& ,
\mu \ll m_i \end{array} \right. \nonumber \\ &&
\end{eqnarray}
where $n$ is the number of external lines
and the upper case applies only to fermions since for $W^\pm$
we have $\mu < M$. Note that in all contributions from the regime $\mu<M$ we have
kept mass terms inside the logarithms. This approach is valid in the entire Standard
Model up to terms of order ${\cal O} \left( \log \frac{m_t}{M} \right)$.

\section{Comparison with one loop results} \label{sec:ol}

In this section we compare our results from the infrared evolution equation method with the explicit
one loop calculation of Ref. \cite{bddms} for longitudinal $W^\pm_{\rm L}$ 
scattering in $e^+e^-$ collisions
and the general one loop results from Ref. \cite{dp}.
In Ref. \cite{dp} all mass singular terms were isolated and the physical basis was used to 
obtain the DL and SL corrections from collinear terms, wave function renormalization and
RG contributions. The results presented there for fermions (up to Yukawa terms) and transverse
degrees of freedom agree with our corresponding results in Ref. \cite{m1}. Also all terms
calculated here are, at one loop, in agreement with Ref. \cite{dp}.
The results of Ref. \cite{bddms} were obtained in terms of the physical fields. 
We already checked that our method gives the
correct terms at one loop for transverse degrees of freedom and for fermions (up to top-Yukawa terms)
in Ref. \cite{m1}. Soft real photon radiation will be included in the comparison.
This comparison is crucial as mentioned in section \ref{sec:sQCD} since we must check that
the splitting function approach, in particular the factorization of the DL and SL terms
takes place with the same electroweak group factor 
$\left( \frac{g^2}{8 \pi^2} T_{\phi} (T_{\phi}+1)
+ \frac{{g^\prime}^2}{8 \pi^2} \frac{Y^2_{\phi}}{4} \right)$ from the high effective scalar
theory.
Only the Yukawa terms factorize differently, namely with $\frac{g^2}{8 \pi^2}$.
In the following, the lower index on the cross section indicates the helicity of the electron, where
$e^-_-$ denotes the left handed electron.
We summarize the relevant
results for 
$e^+_+ e^-_- \longrightarrow W^+_{\rm L} W^-_{\rm L}$ and $e^+_- e^-_+ \longrightarrow 
W^+_{\rm L}
W^-_{\rm L}$ from Ref. \cite{bddms} for convenience as follows:
\begin{eqnarray}
&& \!\!\!\!\!\!\!\!\!\!\! \left( \frac{ d \sigma}{d \Omega} \right)_{-,\,{\rm L}} \!\!\! \approx \!
\left( \frac{d \sigma}{d \Omega
}
\right)^{\rm Born}_{-,\,{\rm L} } \!\! \left\{ \! 1 \! + \frac{e^2}{8\pi^2} \! \left[ - \frac{1-2c_{\rm w}^2+4
c_{\rm w}^4}{
2 c_{\rm w}^2 s_{\rm w}^2} \log^2 \! \frac{s}{M^2}
+ \frac{103-158 c_{\rm w}^2+80 c_{\rm w}^4}{12 c_{\rm w}^2 s_{\rm w}^2} \log \frac{s}{M^2}
\right. \right. \nonumber \\ && \;\;\;\;\;\;\;\;\;\;\; 
- \frac{3 m_t^2}{2 s_{\rm w}^2 M^2} \log \frac{s}{M^2}+3 \log \frac{s}{m_e^2}+ 2 \log \frac{4 \Delta E^2}{s} \left( \log \frac{s}{m_e^2} + \log \frac{s}{
M^2} -2 \right) \nonumber \\ && \;\;\;\;\;\;\;\;\;\;\; \left. \left. - \frac{4}{3} \sum_{j=1}^{n_f} Q_j^2 N^j_C 
\log \frac{m_j^2}{M^2} \right] \right\}
\label{eq:ml} \\
&& \!\!\!\!\!\!\!\!\!\!\! \left( \frac{ d \sigma}{d \Omega} \right)_{+,\,{\rm L}} 
\!\!\! \approx \! \left( \frac{d \sigma}{d \Omega
}
\right)^{\rm Born}_{+,\,{\rm L}} \! \left\{ \! 1 \!+ \frac{e^2}{8\pi^2} \left[ - \frac{5-10c_{\rm w}^2+8
c_{\rm w}^4}{
4 c_{\rm w}^2 s_{\rm w}^2} \log^2 \frac{s}{M^2} + \frac{65-65 c_{\rm w}^2+18 c_{\rm w}^4}{6 c_{\rm 
w}^2 s_{\rm w}^2} \log \frac{s}{M^2}
\right. \right. \nonumber \\ && \;\;\;\;\;\;\;\;\;\;\; 
- \frac{3 m_t^2}{2 s_{\rm w}^2 M^2} \log \frac{s}{M^2}+3 \log \frac{s}{m_e^2}+ 2 \log \frac{4 \Delta E^2}{s} \left( \log \frac{s}{m_e^2} + \log \frac{s}{
M^2} -2 \right) \nonumber \\ && \;\;\;\;\;\;\;\;\;\;\; \left. \left. - \frac{4}{3} \sum_{j=1}^{n_f} Q_j^2 N^j_C 
\log \frac{m_j^2}{M^2} \right] \right\}
\label{eq:pl} 
\end{eqnarray}
The Born cross sections are given by:
\begin{eqnarray}
\left( \frac{d \sigma}{d \Omega} \right)^{\rm Born}_{-,\,{\rm L}} &=& \frac{e^4}{64 \pi^2s}
\frac{1}{16 s_{\rm w}^4 c_{\rm w}^4} \sin^2 \theta \label{eq:bmp} \\
\left( \frac{d \sigma}{d \Omega} \right)^{\rm Born}_{+,\,{\rm L}} &=& \frac{e^4}{64 \pi^2s}
\frac{1}{4 c_{\rm w}^4} \sin^2 \theta \label{eq:bpp}
\end{eqnarray}
These expressions demonstrate that the longitudinal cross sections
in Eqs. (\ref{eq:bmp}) and (\ref{eq:bpp}) are not mass suppressed.
Eqs. (\ref{eq:ml}) and (\ref{eq:pl}) were of course calculated in terms of the physical fields
of the broken theory and in the on-shell scheme. We denote $c_{\rm w}=\cos \theta_{
\rm w}$ and $s_{\rm w}=\sin \theta_{\rm w}$ respectively.
Using $e= \frac{g g^\prime}{\sqrt{g^2+{g^\prime}^2}}$, $s_{\rm w}=
\frac{g^\prime}{\sqrt{g^2+{g^\prime}^2}}$ and $c_{\rm w}= \frac{g}{\sqrt{g^2+
{g^\prime}^2}}$ we see that the Born cross section
in Eq. (\ref{eq:bmp}) is proportional to $(g^2+{g^\prime}^2)^2$ and Eq. (\ref{eq:bmp})
proportional to ${g^\prime}^4$.
Below the scale where non-Abelian effects enter, we have running coupling corrections only
from QED, i.e. $g^2(M^2)=\frac{e_{\rm eff}^2(M^2)}{s^2_{\rm w}}$ and
${g^\prime}^2(M^2)=\frac{e_{\rm eff}^2(M^2)}{c^2_{\rm w}}$ where
\begin{equation}
e^2_{\rm eff}(M^2)=e^2 \left(1+ \frac{1}{3} \frac{e^2}{4 \pi^2} \sum_{j=1}^{n_f} Q_j^2 N^j_C \log 
\frac{M^2}{m_j^2} \right)
\end{equation}
Thus, the RG-corrections to both cross sections for $\overline{\mu}>M$ 
are given by (using $C_A$=2, $n_{gen}=3$ and
$n_h=1$ in Eqs. \ref{eq:ewb0}):
\begin{eqnarray}
&& \left( \frac{d \sigma}{d \Omega} \right)^{\rm RG}_{-,\,{\rm L}} = 
\left( \frac{d \sigma}{d \Omega} \right)^{\rm Born}_{-,\,{\rm L}} \left\{ 1 +
\frac{e^2}{8 \pi^2} \frac{41-82 c^2_{\rm w} + 22 c^4_{\rm w}}{6 s^2_{\rm w}c^2_{\rm w}} \log
\frac{s}{M^2} \right\} \label{eq:mlrg} \\
&& \left( \frac{d \sigma}{d \Omega} \right)^{\rm RG}_{+,\,{\rm L}} =
\left( \frac{d \sigma}{d \Omega} \right)^{\rm Born}_{+,\,{\rm L}} \left\{1+
\frac{e^2}{8 \pi^2} \frac{41}{6 c^2_{\rm w}} \log \frac{s}{M^2} \right\} \label{eq:plrg}
\end{eqnarray}
\begin{table}
\begin{center}
\begin{Large}
\begin{tabular}{|l|c|c|r|}
\hline
& T & Y & Q \\
\hline
$e^-_-$ & 1/2 & -1 & -1 \\
\hline
$e^-_+$ & 0 & -2 & -1 \\
\hline
$e^+_+$ & 1/2 & 1 & 1 \\
\hline
$e^+_-$ & 0 & 2 & 1 \\
\hline
$u_-$ & 1/2 & 1/3 & 2/3 \\
\hline
$u_+$ & 0 & 4/3 & 2/3 \\
\hline
$d_-$ & 1/2 & 1/3 & -1/3 \\
\hline
$d_+$ & 0 & -2/3 & -1/3 \\
\hline
$W^\pm$ & 1 & 0 & $\pm$1 \\
\hline
$\phi^\pm$ & 1/2 & $\pm$1 & $\pm$1 \\
\hline
$\chi$ & 1/2 & $-$1 & 0 \\
\hline
$H$ & 1/2 & $+$1 & 0 \\
\hline
\end{tabular}
\end{Large}
\end{center}
\caption{The quantum numbers of various particles in the electroweak theory.
The indices indicate the helicity of the electrons and quarks. We neglect all mass terms, i.e.
consider all particles as chiral eigenstates with well defined total weak isospin (T)
and weak hypercharge (Y) quantum numbers. In each case, the electric
charge $Q$, measured in units of the proton charge, by the
Gell-Mann-Nishijima formula $Q=T^{3}+Y/2$.
For longitudinally
polarized gauge bosons, the associated scalar Goldstone bosons describe the DL asymptotics.}
\label{tab:qn}
\end{table}
The Sudakov corrections to both cross sections from the infrared evolution equation
method according to Eq. (\ref{eq:si}) in the soft photon approximation are given below. The
quantum numbers are those of the particle-indices and are summarized in Tab. \ref{tab:qn}:
\begin{eqnarray}
\left( \frac{ d \sigma}{d \Omega} \right)_{-,{\rm L}}\!\! &=& \!\! \left( \frac{d \sigma}{d \Omega}
\right)^{\rm Born}_{-,{\rm L}} \left\{ 1 - \left( \frac{g^2}{8 \pi^2} T_{\phi} (T_{\phi}+1)
+ \frac{{g^\prime}^2}{8 \pi^2} \frac{Y^2_{\phi}}{4} \right) \left( \log^2 \frac{s}{M^2} 
-4 \log \frac{s}{M^2} \right) \right. \nonumber
\\
\!\! && \!\! - \left( \frac{g^2}{8 \pi^2} T_{e^-_-} (T_{e^-_-}+1) + \frac{{g^\prime}^2}{8 \pi^2} 
\frac{Y^2_{e^-_-}}{4}
\right) \left( \log^2 \frac{s}{M^2}- 3 \log \frac{s}{M^2} \right) 
\nonumber \\ \!\! && \!\!
- 3 \frac{g^2}{16 \pi^2} \frac{m_t^2}{M^2} \log \frac{s}{M^2} - \frac{e^2}{8 \pi^2}
\left[ \left( \log \frac{s}{m_e^2} - 1 \right) 2 \log \frac{m_e^2}{\mu^2} \right. \nonumber \\
\!\! && \!\! 
+\log^2 \frac{s}{m_e^2}-3 \log \frac{s}{m_e^2}- \log^2 \frac{s}{M^2}+3 \log \frac{s}{M^2}
\nonumber \\ \!\! && \!\! + 2 \left( \log \frac{s}{M^2}-1 \right) \log \frac{M^2}{\mu^2} - \left(
\log \frac{s}{m_e^2} - 1 \right) \left( 2 \log \frac{m_e^2}{\mu^2}-2 \log \frac{s}{\mu^2_{exp}} 
\right)
- \nonumber \\ \!\! && 2 \left( \log \frac{s}{M^2}-1 \right)
\left( \log \frac{M^2}{\mu^2} - \log \frac{s}{\mu^2_{exp}} \right) - \log^2 \frac{s}{m_e^2} -
\log^2 \frac{s}{M^2} \Bigg]
\nonumber \\
\!\! && \left. + \frac{2}{3} \frac{e^2}{4 \pi^2} \sum_{j=1}^{n_f} Q_j^2 N^j_C \log \frac{M^2}{m_j^2}
 \right\}
 \nonumber \\
 \!\! &=& \!\! \left( \frac{d \sigma}{d \Omega}\right)^{\rm \!\! Born}_{-,{\rm L}} \!
 \left\{ 1 - \frac{e^2}{8 \pi^2} \left( \frac{1+2 c_{\rm w}^2}{2 s_{\rm w}^2 c_{\rm w}^2}
 \log^2 \frac{s}{M^2}
 - 7 \frac{1+2 c_{\rm w}^2}{4 s_{\rm w}^2 c_{\rm w}^2}
 \log \frac{s}{M^2} \right) + \frac{e^2}{8 \pi^2} \times \right. \nonumber \\
 \!\! && \!\! \left[ 2 \log^2 \frac{s}{M^2} - 3 \log \frac{m_e^2}{M^2} - 4 \log
 \frac{s}{\mu^2_{exp}} \left( \log \frac{s}{m_e M}-1 \right)-  
 \frac{3 m_t^2}{2 s^2_{\rm w} M^2} \log \frac{s}{M^2} \right] \nonumber \\
 \!\! && \left. + \frac{2}{3} \frac{e^2}{4 \pi^2} \sum_{j=1}^{n_f} Q_j^2 N^j_C \log \frac{M^2}{m_j^2}
  \right\}
\label{eq:myml}
\end{eqnarray}
Adding Eqs. (\ref{eq:mlrg}) and (\ref{eq:myml}) yields exactly the result in Eq. (\ref{eq:ml}) from
Ref. \cite{bddms}. Analogously, we have for right handed electrons:
\begin{eqnarray}
\left( \frac{ d \sigma}{d \Omega} \right)_{+,{\rm L}}\!\! &=& \!\! \left( \frac{d \sigma}{d \Omega}
\right)^{\rm Born}_{+,{\rm L}} \left\{ 1 - \left( \frac{g^2}{8 \pi^2} T_{\phi} (T_{\phi}+1)
+ \frac{{g^\prime}^2}{8 \pi^2} \frac{Y^2_{\phi}}{4} \right) \left( \log^2 \frac{s}{M^2} 
-4 \log \frac{s}{M^2} \right) \right. \nonumber
\\
\!\! && \!\! - \left( \frac{g^2}{8 \pi^2} T_{e^-_+} (T_{e^-_+}+1) + \frac{{g^\prime}^2}{8 \pi^2} 
\frac{Y^2_{e^-_+}}{4}
\right) \left( \log^2 \frac{s}{M^2}- 3 \log \frac{s}{M^2} \right) 
\nonumber \\ \!\! && \!\!
- 3 \frac{g^2}{16 \pi^2} \frac{m_t^2}{M^2} \log \frac{s}{M^2} - \frac{e^2}{8 \pi^2}
\left[ \left( \log \frac{s}{m_e^2} - 1 \right) 2 \log \frac{m_e^2}{\mu^2} \right. \nonumber \\
\!\! && \!\!
+\log^2 \frac{s}{m_e^2}-3 \log \frac{s}{m_e^2}- \log^2 \frac{s}{M^2}+3 \log \frac{s}{M^2}
\nonumber \\ \!\! && \!\! + 2 \left( \log \frac{s}{M^2}-1 \right) \log \frac{M^2}{\mu^2} - \left(
\log \frac{s}{m_e^2} - 1 \right) \left( 2 \log \frac{m_e^2}{\mu^2}-2 \log \frac{s}{\mu^2_{exp}} 
\right)
- \nonumber \\ \!\! && 2 \left( \log \frac{s}{M^2}-1 \right)
\left( \log \frac{M^2}{\mu^2} - \log \frac{s}{\mu^2_{exp}} \right) - \log^2 \frac{s}{m_e^2} -
\log^2 \frac{s}{M^2} \Bigg]
\nonumber \\
\!\! && \left. + \frac{2}{3} \frac{e^2}{4 \pi^2} \sum_{j=1}^{n_f} Q_j^2 N^j_C \log \frac{M^2}{m_j^2}
 \right\}
 \nonumber \\
 \!\! &=& \!\! \left( \frac{d \sigma}{d \Omega}\right)^{\rm \!\! Born}_{+,{\rm L}} \!
 \left\{ 1 - \frac{e^2}{8 \pi^2} \left( \frac{5-2 c_{\rm w}^2}{4 s_{\rm w}^2 c_{\rm w}^2}
 \log^2 \frac{s}{M^2}
 -  \frac{4- c_{\rm w}^2}{s_{\rm w}^2 c_{\rm w}^2}
 \log \frac{s}{M^2} \right) + \frac{e^2}{8 \pi^2} \times \right. \nonumber \\
 \!\! && \!\! \left[ 2 \log^2 \frac{s}{M^2} - 3 \log \frac{m_e^2}{M^2} - 4 \log
 \frac{s}{\mu^2_{exp}} \left( \log \frac{s}{m_e M}-1 \right) -
 \frac{3 m_t^2}{2 s^2_{\rm w} M^2} \log \frac{s}{M^2} \right] \nonumber \\
 \!\! && \left. + \frac{2}{3} \frac{e^2}{4 \pi^2} \sum_{j=1}^{n_f} Q_j^2 N^j_C \log \frac{M^2}{m_j^2}
  \right\}
\label{eq:mypl}
\end{eqnarray}
Again we see that after adding Eqs. (\ref{eq:plrg}) and (\ref{eq:mypl}) we obtain the result in 
Eq. (\ref{eq:pl}) from
Ref. \cite{bddms}. Thus we have demonstrated that to subleading logarithmic accuracy our results
from the infrared evolution equation method in conjunction with the Goldstone boson equivalence
theorem are
identical with existing one loop calculations with physical fields in the high energy limit.

\section{Discussion of the results} \label{sec:dis}

In this section we discuss the size of the subleading Sudakov corrections obtained in this work.
We neglect renormalization group corrections for simplicity and use $\frac{e^2}{4 \pi}=\frac{1}{137}$,
$\frac{g^2}{4 \pi}=\frac{e^2(M^2)}{s^2_{\rm w} 4 \pi} = \frac{1}{0.23 \times 128}$
and $\frac{{g^\prime}^2}{4 \pi}=\frac{e^2(M^2)}{c^2_{\rm w} 4 \pi} = \frac{1}{0.77 \times 128}$.
The motivation for investigating
the size of the gauge invariant corrections at the subleading level 
is two-fold.
While this discussion is incomplete for processes with a large angular dependence, it
is nevertheless useful in estimating how good the DL approximation is at higher orders.
In addition, we gain physical insight into the importance of Yukawa corrections and
the partial cancellation between subleading terms.

\subsection{Sudakov effects for longitudinal gauge boson and Higgs production}

We begin with the Yukawa corrections
for external scalars given in Eq. (\ref{eq:Wphi})
with the infrared cutoff $\mu=M$.
Using the quantum numbers of Tab. \ref{tab:qn}, we have
\begin{eqnarray}
 -W^\phi_i(s,M^2) \!\!\!&=& \!\!\! - \frac{ g^2}{16 \pi^2} \!\! \left[ \! \left( \frac{1}{2} \left(
 \frac{1}{2}+1 \! \right)+  \tan^2 \! 
 \theta_{\rm w}  
 \frac{1}{4} \right) \!\! \left( \log^2 \frac{s}{M^2}- 4 \log \frac{s}{M^2} 
 \! \right) \!\!
 + \frac{3}{2} \frac{m^2_t}{M^2} \log \frac{s}{M^2} \right]  \nonumber \\
 &=& \!\!\! -\frac{ g^2}{16 \pi^2} \!\! \left[ \! 0.79 \log^2 \frac{s}{M^2} + 4.01 \log \frac{s}{M^2}
 \right]
 \label{eq:Wphinum}
\end{eqnarray}
where we use $M=80$ GeV, $m_t=175$ GeV and $s^2_{\rm w}=0.23$. 
The first thing to notice is that the Yukawa enhanced logarithms
dominate over the subleading Sudakov corrections and enhance the overall
Sudakov suppression. At 1 TeV we have $\log \frac{s}{M^2}=5.05$ and thus almost equal contributions
from DL and SL terms. At 2 TeV we have $\log \frac{s}{M^2}=6.44$ and at 3 TeV 
$\log \frac{s}{M^2}=7.25$.
In real calculations, however, one finds that the Yukawa terms are always proportional to $\log 
\frac{s}{m_t^2}$. Since the factor of the Yukawa logarithm is uniquely determined
by Eq. (\ref{eq:Wphi}) we can replace the respective mass term inside the
logarithm\footnote{Analogously for $\chi$ we can put $M=M_{\rm Z}$ and for
$H$ we have $M=M_{\rm H}$ as arguments of the non-Yukawa logarithms in Eq. (\ref{eq:Wphi})
depending on which mass is the largest in a given process.} 
Thus, for $\phi^\pm$ for instance, we have to consider
\begin{eqnarray}
 -W^\phi_i(s,M^2) \!\!\!&=& \!\!\! - \frac{ g^2}{16 \pi^2} \!\! \left[ \! \left( \frac{1}{2} \left(
 \frac{1}{2}+1 \right)+  \tan^2 \! 
 \theta_{\rm w}  
 \frac{1}{4} \right) \!\! \left( \log^2 \frac{s}{M^2}- 4 \log \frac{s}{M^2} 
 \! \right) \!\!
 + \frac{3}{2} \frac{m^2_t}{M^2} \log \frac{s}{m_t^2} \right]  \nonumber \\
 &\approx& \!\!\! - \frac{ g^2}{16 \pi^2} \!\! \left[ \! 0.79 \log^2 \frac{s}{M^2} 
 + 4.01 \log \frac{s}{M^2} - 11.24
 \right]
 \label{eq:Wphimt}
\end{eqnarray}
With the above mass values
we have at one loop about 40 (39) \% at 1 (3) TeV from the subleading terms
relative to the DL corrections and 
at the two loop level about 79 (77) \% at 1 (3) TeV.
The subleading corrections are therefore non-negligible and
enhancing the Sudakov suppression. Even at 100 TeV the subleading terms make up about
52 \% (!) at the two loop level and must be taken into account. The good news is that
the absolute size of the DL correction per line according
to Eq. (\ref{eq:Wphimt}) at the one loop level is 5.7 (11.7) \% at 1 (3) TeV and
at two loops 0.16 (0.7) \% at 1 (3) TeV relative to the Born cross section. 

The above numbers are valid for both external $H$ and $\chi$ fields. For $\phi^\pm$ we
also have to consider the purely electromagnetic corrections according to Eq. (\ref{eq:si}).
Thus we have on the level of the cross section for each longitudinally polarized $W$-boson
including soft photon radiation:
\begin{eqnarray}
&& -w_i^{\rm w} (s, \mu^2) + w_i^{\rm w} (s, M^2) + w_{\rm expt}^\gamma (s,M,\mu,\mu_{\rm expt})
= \nonumber \\
&& \frac{e^2}{16 \pi^2} \left[ \log^2 \frac{s}{M^2} - 2 \log \frac{s}{\mu^2_{\rm expt}}
\left( \log \frac{s}{M^2} - 1 \right) \right]
\end{eqnarray}
Thus, the complete size of the corrections for $\phi^\pm$ on the level of the cross section, 
chosing $\mu_{\rm expt}=M$, is given by
\begin{eqnarray}
&& - W^\phi_i(s,M^2) -w_i^{\rm w} (s, \mu^2) + w_i^{\rm w} (s, M^2) + 
w_{\rm expt}^\gamma (s,M,\mu,M) = \nonumber \\
&& - \frac{ g^2}{16 \pi^2} \!\! \left[ \! 1.02 \log^2 \frac{s}{M^2} 
 + 2.01 \log \frac{s}{M^2} - 11.24
 \right]
 \label{eq:Wphicomp}
\end{eqnarray}
It is clear that the DL approximation is much more appropriate for longitudinal $W$-bosons
than for the neutral external scalars. For instance we have about 
6.5 (4.3) \% at 1 (3) TeV from the subleading terms relative to the DL contributions.
The absolute size of the DL corrections relative to the Born cross section at one loop 
is 7 (15) \% at 1 (3) TeV and at two loops 0.26 (1.1) \% at 1 (3) TeV. 
The subleading terms at the two loop level contribute about 13 (8.6) \% relative to the
DL corrections.

\subsection{Sudakov effects for quarks of the third generation}

In order to estimate the size of the corrections for chiral heavy quark
production we consider first the case of left handed
bottom and top quarks. In this case we have from Tab. \ref{tab:qn}:
\begin{eqnarray}
 -W_i^{t_{\rm L},b_{\rm L}}(s,M^2) \!\!\!&=& \!\!\!\!- \frac{ g^2}{16 \pi^2} \!\! \left[ \! 
 \left( \frac{1}{2} \! \left(\frac{1}{2} \!+ \!1 \!\right)
 \! + \tan^2 \! 
 \theta_{\rm w}  
 \frac{1}{36} \right) \!\! 
\left( \log^2 \!\! \frac{s}{M^2}- 3 \log \frac{s}{M^2} 
 \! \right) \!\!+ \!
 \frac{m^2_t}{4 M^2}  
 \log \frac{s}{m_t^2} \right] \nonumber \\
 &\approx& \!\!\!\! -\frac{ g^2}{16 \pi^2} \!\! \left[ \! 0.814 \log^2 \frac{s}{M^2}
 -1.246 \log \frac{s}{M^2} - 1.87 \right]
 \label{eq:Wfmt}
\end{eqnarray}
The corrections of purely electromagnetic origin are different for the two cases.
In general we have
\begin{eqnarray}
&& -w_i^{f} (s, \mu^2) + w_i^{f} (s, M^2) + w_{\rm expt}^\gamma (s,m_f,\mu,M)
= \nonumber \\
&& \frac{e^2Q^2_f}{16 \pi^2} \left[ - \log^2 \frac{s}{M^2} - \log \frac{M^2}{m_f^2}
\left( 2 \log \frac{s}{M^2} - 3 \right)  
+ 2 \log \frac{s}{M^2} \right] \label{eq:emf}
\end{eqnarray}
The full result for the left handed top quark is therefore given by: 
\begin{eqnarray}
&& -W_i^{t_{\rm L}}(s,M^2) -w_i^{t_{\rm L}} (s, \mu^2) + w_i^{t_{\rm L} } (s, M^2) + 
w_{\rm expt}^\gamma (s,m_t,\mu,M)
= \nonumber \\
&& - \frac{ g^2}{16 \pi^2} \!\! \left[ \! 0.916 \log^2 \frac{s}{M^2}
 -1.769 \log \frac{s}{M^2} - 1.39 \right]
 \label{eq:Wtmtcomp}
\end{eqnarray}
Thus we see that there is a partial cancellation between the subleading and the Yukawa terms
and the overall DL suppression is somewhat reduced. In relative terms at one loop, the
SL corrections are about 33 (20) \% at 1 (3) TeV, and at two loops the relative size of
the subleading terms is 65 (41) \% at 1 (3) TeV. The absolute size of the DL corrections
at one loop is 5.4 (11.1) \% per line at 1 (3) TeV. At two loops we have corrections
of 0.15 (0.62) \% at 1 (3) TeV relative to the Born cross section.

The full result for the left handed bottom quark (with $m_b=4.5$ GeV) is given by: 
\begin{eqnarray}
&& -W_i^{b_{\rm L}}(s,M^2) -w_i^{b_{\rm L}} (s, \mu^2) + w_i^{b_{\rm L} } (s, M^2) + 
w_{\rm expt}^\gamma (s,m_b,\mu,M)
= \nonumber \\
&& - \frac{ g^2}{16 \pi^2} \!\! \left[ \! 0.829 \log^2 \frac{s}{M^2}
 -1.002 \log \frac{s}{M^2} - 2.31 \right]
 \label{eq:Wbmtcomp}
\end{eqnarray}
The partial cancellation between the subleading and the Yukawa terms
and the overall DL suppression is reduced. In relative terms at one loop, the
SL corrections are about 43 (29) \% at 1 (3) TeV, and at two loops the relative size of
the subleading terms is 86 (58) \% at 1 (3) TeV. The absolute size of the DL corrections
at one loop is 5.9 (12.2) \% per line at 1 (3) TeV. At two loops we have corrections
of 0.18 (0.75) \% at 1 (3) TeV relative to the Born cross section.

For a right handed 
top quark we have from Tab. \ref{tab:qn}:
\begin{eqnarray}
- W_i^{t_{\rm R}}(s,M^2) \!\!&=& \!\!- \frac{ g^2}{16 \pi^2} \!\! \left[ \! 
 \frac{4}{9} \frac{s^2_{\rm w}}{c_{\rm w}^2}  
\left( \log^2\frac{s}{M^2} - 3 \log \frac{s}{M^2} 
 \! \right) 
 +\frac{m^2_t}{2 M^2}  
 \log \frac{s}{m_t^2} \right] \nonumber \\
 &\approx& \!\! -\frac{ g^2}{16 \pi^2} \!\! \left[ \! 0.133 \log^2 \frac{s}{M^2}
 +2.26 \log \frac{s}{M^2} - 3.746 \right]
 \label{eq:Wfmr}
\end{eqnarray}
Now we need to add the corrections from Eq. (\ref{eq:emf}). The full result is thus
\begin{eqnarray}
&& -W_i^{t_{\rm R}}(s,M^2) -w_i^{t_{\rm R}} (s, \mu^2) + w_i^{t_{\rm R} } (s, M^2) + 
w_{\rm expt}^\gamma (s,m_t,\mu,M)
= \nonumber \\
&& - \frac{ g^2}{16 \pi^2} \!\! \left[ \! 0.235 \log^2 \frac{s}{M^2}
 +1.737 \log \frac{s}{M^2} - 3.27 \right]
 \label{eq:Wtmtcompr}
\end{eqnarray}
Thus we see that there is a large correction of the top Yukawa terms (a factor of 7.4 
for the relative coefficients)
and the overall DL suppression is strongly enhanced. In relation to the DL contribution
at one loop, the
SL corrections are about 69 (60) \% at 1 (3) TeV, and at two loops the relative size of
the subleading terms is 139 (120) \% at 1 (3) TeV. The absolute size of the DL corrections
at one loop is 1.6 (3.3) \% per line at 1 (3) TeV. At two loops, however, we have corrections
of only 0.013 (0.055) \% at 1 (3) TeV relative to the Born cross section. Thus, the apparent
lack of convergence of the DL approximation is irrelevant for practical purposes.

From a physical point of view it is clear how the  large subleading terms can be understood.
Right handed fermions in general couple only to photons and Z-bosons and 
the coupling to the Z-boson is proportional to $\left(
\frac{\alpha^\prime}{4 \pi} \right)$ for
the DL and non-Yukawa SL corrections. The Yukawa corrections, however are proportional to
$\left( \frac{\alpha}{4 \pi} \right)$ and in addition for the right handed top, we have:
\begin{eqnarray}
\frac{m_t^2}{2M^2} &\sim & 5.38 \frac{Y_t^2}{4}= 5.38 \frac{4}{9} \\  
\alpha (M^2) &\sim & 3.35 \alpha^\prime (M^2)
\end{eqnarray}
The effect is somewhat softened by the electromagnetic corrections from Eq. (\ref{eq:emf}).
In general the size of the subleading terms cannot be neglected at the two loop level for
all Yukawa enhanced Sudakov corrections discussed in this work.

\section{Conclusions} \label{sec:co}

In this paper we calculated the universal subleading ${\cal O} \left( 
g^{2n} \log^{2n-1} \frac{s}{M^2} , g^{\prime^{2n}} \log^{2n-1} \frac{s}{M^2}
\right)$ logarithmic Sudakov corrections to longitudinal gauge boson and Higgs production
to all orders. We have employed the infrared evolution equation method and used the equivalence
theorem to obtain the high energy kernel of the equation for longitudinal gauge boson 
production. All Yukawa enhanced SL Sudakov terms in non-mass suppressed amplitudes
are universal to all orders. This feature is evident in the splitting function formalism
which we have adopted to calculate the virtual Sudakov corrections.
The approach, concerning in particular the novel Yukawa enhanced subleading corrections,
has been verified by employing 
a non-Abelian generalization of Gribov's bremsstrahlung theorem. 
We agree with the literature at the one loop level, which is a highly non-trivial
check considering the complicated nature of electroweak radiative corrections
and can serve as an independent confirmation of those results.
In addition this comparison confirms the validity of the splitting function approach since
DL and non-Yukawa SL corrections factorize with respect to the same group factor of the effective high
energy theory. These SL contributions are determined by the spin only and are thus
identical to those found in a scalar theory with an unbroken $SU(2)\times U(1)$, while the
Yukawa enhanced SL corrections indicate the spontaneously broken
gauge symmetry.

The physical picture which is now emerging is clear: at high energies the
SM behaves like an unbroken gauge theory up to DL and SL accuracy for fermions and 
transversely polarized gauge bosons. Only Yukawa corrections are novel features in this
picture. For longitudinally polarized gauge bosons and Higgs scalars, the effective theory
is given by the Goldstone boson equivalence theorem and contains corrections in analogy to
a non-Abelian gauge theory with scalar fields in the fundamental representation. Again,
Yukawa terms modify this picture as a unique ingredient of broken gauge theories.
The mass gap between the electroweak gauge bosons can be included in a natural way via
the matching conditions in the framework of the infrared evolution equation method.
Thus all universal Sudakov corrections to DL and SL accuracy are known in the electroweak
theory to all orders. 
The remaining corrections which enter at this level of precision
are given by angular terms of the type $\log \frac{u}{t} \log \frac{s}{M^2}$. 
These terms are non-universal and don't factorize with respect to the Born amplitude.
While these
terms are known at one loop, for phenomenological applications at future colliders a
two loop analysis is desirable.
In addition, subleading RG effects of the type $\alpha^n \beta_0 \log^{2n-1} \frac{s}{M^2}$ 
coupling effects at higher order should be consistently resummed 
via the inclusion of a running coupling in each loop analogously to the QCD Sudakov form factor.

In summary, all universal Sudakov logarithms in the electroweak SM are known at the subleading level
to all orders and are non-negligible at future collider energies.
The inclusion of the leading and full subleading electroweak radiative corrections at least
at the two loop level will be important in investigating new physics effects
at TeV energies. 

\section*{Acknowledgments} 

We would like to thank J.~Collins, A.~Denner and S.~Pozzorini for discussions.

\end{document}